\documentclass[twocolumn]{aastex63}
\received{June 1, 2019}
\revised{January 10, 2019}
\accepted{\today}
\submitjournal{AJ}

\shorttitle{The misfired outburst of Cen X-4}
\shortauthors{Baglio M. C. et al. }
\graphicspath{{./}{figures/}}

\begin{document}

\title{A misfired outburst in the neutron star X-ray binary Centaurus X-4}

\correspondingauthor{M. C. Baglio}
\email{mcb19@nyu.edu}

\author[0000-0003-1285-4057]{M. C. Baglio}
\affiliation{Center for Astro, Particle and Planetary Physics, New York University Abu Dhabi, PO Box 129188, Abu Dhabi, UAE \\}
\affiliation{New York University Abu Dhabi, PO Box 129188, Abu Dhabi, United Arab Emirates}
\affiliation{INAF, Osservatorio Astronomico di Brera, Via E. Bianchi 46, I-23807 Merate (LC), Italy}

\author[0000-0002-5319-6620]{P. Saikia}
\affiliation{Center for Astro, Particle and Planetary Physics, New York University Abu Dhabi, PO Box 129188, Abu Dhabi, UAE \\}
\affiliation{New York University Abu Dhabi, PO Box 129188, Abu Dhabi, United Arab Emirates}

\author[0000-0002-3500-631X]{D. M. Russell}
\affiliation{Center for Astro, Particle and Planetary Physics, New York University Abu Dhabi, PO Box 129188, Abu Dhabi, UAE \\}
\affiliation{New York University Abu Dhabi, PO Box 129188, Abu Dhabi, United Arab Emirates}

\author{J. Homan}
\affiliation{Eureka Scientific, Inc., 2452 Delmer Street, Oakland, CA 94602, USA\\}

\author[0000-0002-5542-8624]{S. Waterval}
\affiliation{Center for Astro, Particle and Planetary Physics, New York University Abu Dhabi, PO Box 129188, Abu Dhabi, UAE \\}
\affiliation{New York University Abu Dhabi, PO Box 129188, Abu Dhabi, United Arab Emirates}

\author{D. M. Bramich}
\affiliation{Center for Astro, Particle and Planetary Physics, New York University Abu Dhabi, PO Box 129188, Abu Dhabi, UAE \\}
\affiliation{Division of Engineering, New York University Abu Dhabi, P.O. Box 129188, Saadiyat Island, Abu Dhabi, UAE}

\author[0000-0001-6278-1576]{S. Campana}
\affiliation{INAF, Osservatorio Astronomico di Brera, Via E. Bianchi 46, I-23807 Merate (LC), Italy}

\author[0000-0003-3352-2334]{F. Lewis}
\affiliation{Faulkes Telescope Project, School of Physics and Astronomy, Cardiff University, The Parade, Cardiff, CF24 3AA, Wales, UK}
\affiliation{Astrophysics Research Institute, Liverpool John Moores University, 146 Brownlow Hill, Liverpool L3 5RF, UK}


\author[0000-0002-5686-0611]{J. Van den Eijnden}
\affiliation{Department of Physics, Astrophysics, University of Oxford, Denys Wilkinson Building, Keble Road, Oxford OX1 3RH, UK}

\author[0000-0003-0168-9906]{K. Alabarta}
\affiliation{Center for Astro, Particle and Planetary Physics, New York University Abu Dhabi, PO Box 129188, Abu Dhabi, UAE \\}
\affiliation{New York University Abu Dhabi, PO Box 129188, Abu Dhabi, United Arab Emirates}
\affiliation{Kapteyn Astronomical Institute, University of Groningen, PO Box 800, NL-9700 AV Groningen, the Netherlands \\}
\affiliation{School of Physics and Astronomy, University of Southampton, Southampton, SO17 1BJ, UK \\}

\author[0000-0001-9078-5507]{S. Covino}
\affiliation{INAF, Osservatorio Astronomico di Brera, Via E. Bianchi 46, I-23807 Merate (LC), Italy}

\author{P. D'Avanzo}
\affiliation{INAF, Osservatorio Astronomico di Brera, Via E. Bianchi 46, I-23807 Merate (LC), Italy}
\author{P. Goldoni}
\affiliation{APC, Astroparticule et Cosmologie, Universit\'{e} Paris Diderot, CNRS/IN2P3, CEA/Irfu, Observatoire de Paris, Sorbonne Paris Cit\'{e}, 10, Rue Alice Domon et L\'{e}onie Duquet, F-75006 Paris, France}

\author[0000-0001-9487-7740]{N. Masetti}
\affiliation{INAF - Osservatorio di Astrofisica e Scienza dello Spazio, via Piero Gobetti 93/3, I-40129 Bologna, Italy}
\affiliation{Departamento de Ciencias F\`{i}sicas, Universidad Andr\'{e}s Bello, Av. Fern\'{a}ndez Concha 700, 7591538 Las Condes, Santiago, Chile}

\author[0000-0002-3348-4035]{T.~Mu\~noz-Darias}
\affiliation{Instituto de Astrof\'{i}sica de Canarias, 38205 La Laguna, Tenerife, Spain}
\affiliation{Departamento de Astrof\'{i}sica, Universidad de La Laguna, E-38206 La Laguna, Tenerife, Spain}



\begin{abstract}
 We report on a long-term optical monitoring of the neutron star X-ray binary Centaurus X-4
performed during the last 13.5 years. This source has been in quiescence since its outburst in
1979. Our monitoring reveals the overall evolution of the accretion disc; we detect short-duration flares, likely originating also in the disc, superimposed with a small-amplitude ($<$0.1
mag) ellipsoidal modulation from the companion star due to geometrical effects. A long-term ($\sim$2300 days) downward
trend, followed by a shorter ($\sim$1000 days) upward one, is observed in the disc light
curve. Such a rise in the optical has been observed for other X-ray binaries preceding outbursts, as predicted by the disc instability model. For Cen X-4, the rise of the optical flux
proceeded for $\sim$3 years, and culminated in a flux increase at all wavelengths (optical-UV-X-rays) at the end of 2020. This increase faded after $\sim$2 weeks, without giving rise
to a full outburst. We suggest that the propagation of an inside-out heating front was ignited due to a partial ionization of hydrogen in the inner disc. The propagation
might have stalled soon after the ignition due to the increasing surface density in the disc that the front encountered while propagating outwards. The stall was likely eased by the low level irradiation of the
outer regions of the large accretion disc, as shown by the slope of the optical/X-ray correlation, suggesting that irradiation does not play a strong role in the optical, compared to other sources of emission.

\end{abstract}



\section{Introduction} \label{sec:intro}
Low mass X-ray binaries (LMXBs) are binary systems hosting a compact object, that can be a neutron star (NS) or a stellar-mass black hole (BH), and a low-mass companion star (with mass $\lesssim 1M_{\odot}$). The latter is typically a main-sequence star, filling its Roche lobe and transferring matter and angular momentum to the compact object through the formation of an accretion disk.
LMXBs can be transient, displaying short and sudden outbursts, with X-ray luminosities that can reach $L_{X}\sim 10^{36}-10^{38}\rm \, erg/s$ and high accretion rates, and longer, quieter intervals of quiescence, with a drop of the X-ray luminosity by up to seven orders of magnitude. At X-ray frequencies, outbursts are typically characterised by a sharp increase of the flux, lasting days--months, and a longer, slower decay, that can take place over weeks or months, until reaching its former quiescent level \citep{Frank1987}.

X-ray radiation typically comes from the internal part of the accretion disc, close to the compact object \citep{Lasota2001}, from the corona (which is a region of hot electron plasma that is thought to surround the compact object and, according to some models, the accretion disc) and, in case of NSs, from the compact object itself; optical radiation, on the other hand, is thought to primarily come from the companion star and the external part of the disc, the latter being dominant during outbursts, plus a contribution in some systems from synchrotron radiation from compact, collimated jets (see e.g. \citealt{Homan2005}; \citealt{Russell2007}; \citealt{Buxton2012}; \citealt{Kalemci2013}; \citealt{Baglio2018}; \citealt{Baglio2020}). A rise in the optical flux is expected to occur as the temperature in the disc increases, triggering the ionization of hydrogen which may start the outburst (see \citealt{Lasota2001} for a review).

The mechanism that triggers such outbursts is still uncertain. The most accredited scenario is called the \textit{disc-instability model} (DIM; see \citealt{Lasota2001}, \citealt{Hameury2020} for reviews).
The DIM was first suggested to explain the outbursts in dwarf novae (a subclass of cataclysmic variables, that display recurrent outbursts; see \citealt{Cannizzo1982}), and then extended to LMXBs due to the analogy that was observed between the two classes of systems during outbursts, in particular regarding their fast-rise and exponential decay (\citealt{vanparadijs1984}; \citealt{Cannizzo1985}). According to the DIM, the instability is driven by the ionization state of hydrogen in the disc. If all the hydrogen in the disc is ionized, the system is considered as stable, as it happens e.g. in persistent LMXBs or in nova-like systems (i.e. the class of cataclysmic variables that show a persistent behaviour). However, if the mass accretion rate or the temperature becomes low enough to allow for the recombination of hydrogen, then a thermal-viscous instability can occur in the disc, that oscillates between a hot, ionized state, that we call outbursts, and a cold, recombined state, that is quiescence. 
When the system is in quiescence, the cold accretion disc accumulates mass until a critical density, and at the same time the temperature rises until the hydrogen ionization temperature is reached at a certain radius (ignition point). 
At the ignition point, two heating fronts are generated (\citealt{Smak1984}; \citealt{Menou1999}), one propagating inwards, and the other outwards. 

Two different types of outbursts can be observed, depending, above all, on how fast the two fronts propagate. ``Inside-out'' outbursts start at small radii, and the inward heating front will fast reach the inner accretion disc; ``outside-in'' outbursts instead are ignited further away in the disc, therefore the propagation towards the inner disc takes longer. 
In addition, in inside-out outbursts the outward heating front propagates towards regions of higher densities, while outside-in fronts will always encounter regions with decreasing surface density \citep{Dubus2001}. Therefore, it is easier for an inside-out outburst to stall, and to develop a cooling front that switches off the outburst. Inside-out outbursts therefore typically propagate slowly, leading to long rise times of the outburst. 

Once the outburst is triggered, accretion continues at high rates, giving rise to the observed high X-ray luminosity. Then the outburst starts to decay, and the disk is depleted, bringing the system back to its quiescent state \citep{Lasota2001}.

This picture is very simplified and many studies have shown that the effect of direct and indirect irradiation from the compact object, plus the evaporation of the accretion disc in a region that is close to the compact object (for example, the hot inner flow, or corona), plus geometrical effects are important to take into account in order for the DIM to work for LMXBs (see \citealt{Dubus1999}, \citealt{Dubus2001}). In particular, irradiation has been found to ease the propagation of the outwards heating front in inside-out outbursts by reducing the critical density needed for a certain ring of the disc to become thermally unstable \citep[][]{Dubus2001}. Moreover, some variations are observed for different systems; for example, the time delay between the occurrence of the disk instability (happening with the beginning of the heating front propagation in the disk), and the actual start of the outburst (when accretion onto the compact object is detected as an increase in X-ray luminosity) can be different from system to system.

Observations of the optical rise to outburst are crucial to probe the DIM (in particular, the measurement of the optical to X-ray delay of the rise to outburst, and the gradual optical long-term increase that is sometimes observed before an outburst is triggered). Unfortunately, such observations are often difficult, since outbursts typically rise in a few days, and are frequently detected only when the X-ray flux rises above the all-sky monitors' detection threshold, therefore missing the initial stages of the optical rise. Such optical to X-ray delays during the rise of an outburst have been measured using optical monitoring and X-ray all-sky monitors in a few systems, such as V404 Cyg ($< $7 days; \citealt{Bernardini2016_precursor}), GRO J1655-40 ($<6$ days; \citealt{Orosz1997}; \citealt{Hameury1997}), XTE J1550-564 ($<9$ days; \citealt{Jain2001}), XTE J1118+480 ($<$10 days; \citealt{Wren2001}; \citealt{Zurita2006}), 4U 1543-47 ($<5$ days), ASASSN-18ey (MAXI J1820+070; $<7$ days; \citealt{Tucker2018}), Aql X-1 (3-8 days; \citealt{Shahbaz1998}; \citealt{Russell2019}), etc. Recently, a delay of 12 days was measured for the NS LMXB SAX J1808.4-3658 \citep[][]{Goodwin2020} using an X-ray instrument more sensitive than an all-sky monitor ($NICER$), giving an important confirmation to the optical to X-ray delay during the onset of outbursts in LMXBs.

It is clear that the continuous optical monitoring of LMXBs is essential in order to obtain such measurements, together with many other possible achievements (like, e.g., the study of the quiescent behaviour of the sources, or the monitoring of the different stages of a LMXB outburst; \citealt{Russell2019}). As part of this effort, we have been monitoring $\sim 50$ LMXBs with the Las Cumbres Observatory (LCO) and Faulkes 2-m and 1-m robotic telescopes since 2008 \citep[][]{Lewis2008}, and recently we developed a pipeline, the X-ray Binary New Early Warning System (XB-NEWS) that is able to process all the collected data as soon as they are acquired, and produces real-time light curves of all the monitored objects \citep[for more details on the project, see][]{Russell2019}. Since the monitoring was started and the pipeline has been routinely running, we have been able to detect the onset of outbursts in a few cases before the X-ray all sky monitors could, like in the case of SAX J1808.4-3658 \citep[][]{Goodwin2020} and the one presented in this work.


\section{Centaurus X-4}
Cen X-4 (short for Centaurus X-4) is a NS LMXB, discovered in July 1969 during an outburst by the X-ray satellite Vela 5B \citep{Conner1969}. The source had a second outburst ten years later, in 1979, as detected by the All-Sky Monitor experiment on the Ariel 5 satellite \citep{Kaluzienski1980}, and radio detections were reported \citep{Hjellming1979}. The optical counterpart was identified with a bright, blue object, which brightened to a magnitude of $V\sim 12.8$ mag from $V\sim 18.7$ mag \citep{Canizares1980}. Later, the companion star was classified as a $0.35\,M_{\odot}$ K5--7 V star, filling a $0.6\, R_{\odot}$ Roche lobe (\citealt{Shahbaz1993}; \citealt{Torres2002}; \citealt{Davanzo2005}; \citealt{Shahbaz2014}). The ratio between the masses of the two stars has also been carefully evaluated by \citet{Shahbaz2014}, thanks to which a relatively accurate estimate of the neutron star mass has been derived ($M_{\rm NS}=1.94^{+0.37}_{-0.85}$).
The orbital period has been measured with different techniques, leading to a period of $\sim 15.1$hr (see \citealt{McClintock1990}; \citealt{Torres2002}; \citealt{Casares2007}).
Cen X-4 is one of the brightest quiescent NS-LMXBs in the optical, with $V\sim 18.7$ mag, and a non-negligible accretion disc contribution at optical frequencies also in quiescence 
(\citealt{Shahbaz1993}, \citealt{Torres2002}, \citealt{Davanzo2005}).
The interstellar absorption is low, $A_{\rm V}=0.31\pm0.16$ mag \citep{Russell2006}, and the distance to the system is $1.2\pm0.2$ kpc \citep{Chevalier1989}, that is reasonably consistent with the most recent estimate obtained with Gaia ($2.1^{+1.2}_{-0.6}$ kpc; \citealt{Bailer2018}).

Cen X-4 has been in quiescence since the end of its second outburst in 1979. In December 2020, signs of a possible gradual brightening over the previous $\sim 3$ years were reported thanks to an optical monitoring of the source performed with the LCO 2-m and 1-m robotic telescopes \citep{Waterval2020}. After 2020 August 31 (MJD 59092), the source was Sun-constrained until 2020 December 30 (MJD 59213); the first LCO observation after the Sun constraint ended showed a significant brightening in all optical bands \citep{Saikia2021}, which then resulted in prominent flaring activity that lasted for $\sim 2$ weeks. By mid-January, the source was back to its quiescent levels at all wavelengths \citep{vandenEijnde2021}. 
In this paper, we present long-term optical monitoring of Cen X--4, which led to the prediction of a possible new outburst, and we report on the subsequent observed flaring activity using optical and X-ray observations. 
For the whole study presented in this work, the following Python packages have been used for coding purposes: Matplotlib \citep[][]{Hunter2007} and NumPy \citep[][]{Vanderwalt2011}. Additional data analyses were done using IDL version 8.7.3.
\section{Observations and data analysis} \label{sec:obs}
\subsection{Optical monitoring with LCO}
Cen X-4 has been regularly monitored in the optical during the last $\sim 13.5$ years with the LCO 2-m and 1-m robotic telescopes, from February 14, 2008 (MJD 54510) to June 30, 2021 (MJD 59395), mostly using $V$, $R$ and $i'$ filters (Tab. \ref{tab:wavelengths}). In total, the monitoring campaign until 2021 June 30 has acquired 316, 183, 315 images in $V$, $R$ and $i'$, respectively, plus 110 and 36 images in the $g'$ and $r'$ filters, respectively. The images have been processed and analysed by the recently developed XB-NEWS pipeline, that downloads the reduced images (i.e. bias, dark, and flat-field corrected images) from the LCO archive\footnote{\url{https://archive.lco.global}}, automatically rejects poor quality reduced images, performs astrometry using Gaia DR2 positions\footnote{\url{https://www.cosmos.esa.int/web/gaia/dr2}}, carries out multi-aperture photometry (MAP; \citealt{Stetson1990}), solves for photometric zero-point offsets between epochs \citep[][]{Bramich2012}, and flux-calibrates the photometry using the ATLAS-REFCAT2 catalog \citep[][]{Tonry2018}. If the target is not detected in an image above the detection threshold, then XB-NEWS performs forced MAP at the target coordinates. In this case, we reject all forced MAP magnitudes with an uncertainty $> 0.25$ mag, as these are very uncertain photometric measurements. The pipeline produces near real-time calibrated light curves. For further details on XB-NEWS, see \citet{Russell2019} and \citet{Goodwin2020}.

\begin{table}[]
    \centering
    \begin{tabular}{cc|cc}
    \hline
         Filter & $\nu_c$ (Hz) & Filter & $\nu_c$ (Hz) \\
         \hline
         $uvw2$ & $1.556\times10^{15}$  & $R$    & $4.680\times10^{14}$ \\
         $uvm2$ & $1.334\times10^{15}$  & $i'$ & $3.979\times10^{14}$  \\
         $uvw1$ & $1.154\times10^{15}$  & $z'$ & $3.286\times10^{14}$  \\
         $u$    & $8.658\times 10^{14}$ & $J $ & $2.419\times10^{14}$   \\
         $g'$   & $6.289\times10^{14}$  &  $H$  & $1.807\times10^{14}$  \\
         $V$    & $5.505\times10^{14}$  &   $K$& $1.389\times10^{14}$\\
         $r'$   & $4.831\times10^{14}$  &  & \\
          \hline
    \end{tabular}
    \caption{Central frequency $\nu_c$ of the UV/optical/NIR filters that are relevant for this work.}
    \label{tab:wavelengths}
\end{table}

By visual inspection of the light curves, the presence of a number of outliers was evident. We therefore performed a systematic search for outliers in the light curves by plotting each band against the other, using observations taken a maximum of 0.5 days apart. We then selected all points lying outside the 2-sigma interval and investigated the corresponding images. The majority of these images (a total of 9 in $V$, $R$ and $i'$-band, respectively) were found to be of poor quality for various reasons (i.e. background issues) and were therefore rejected.

\begin{figure*}
    \centering
    \includegraphics[width=19cm]{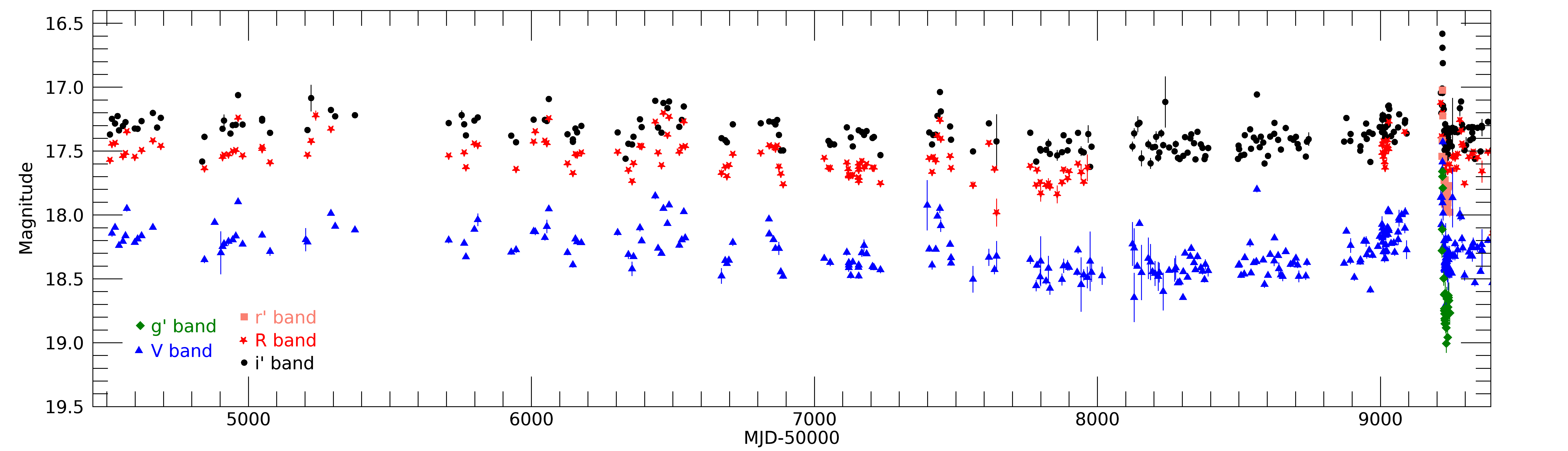}
    \caption{13.5 years of optical monitoring of Cen X-4 performed with LCO in $g'$, $V$, $r'$, $R$ and $i'$ bands. All magnitudes are calibrated; error bars represent $1\sigma$ uncertainties.}
    \label{fig:long_monitoring}
\end{figure*}

\begin{figure}
   \centering
  \includegraphics[width=8.5cm]{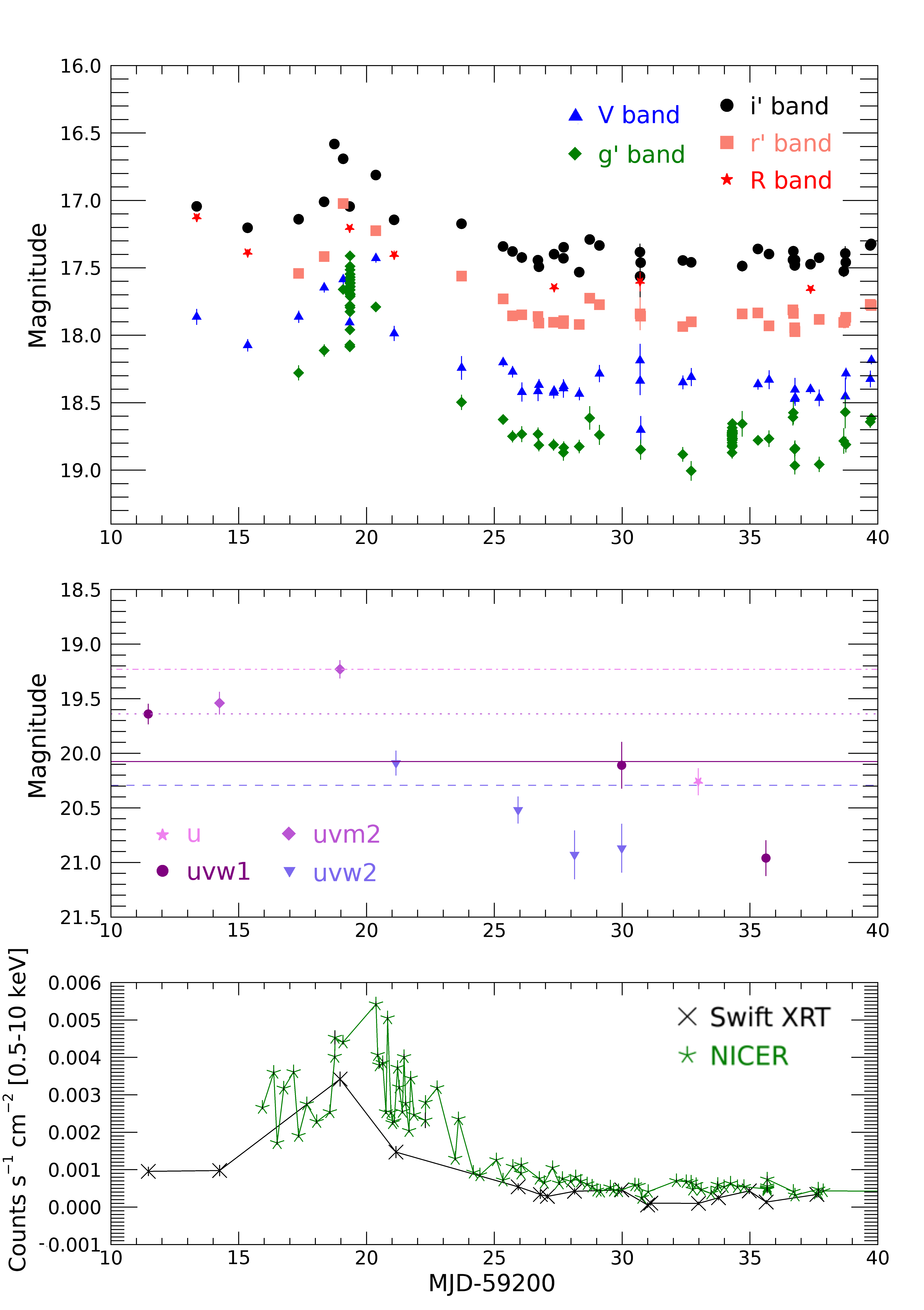}
   \caption{\textit{Top} panel: zoom of the LCO light curve between 2020, December 30 (MJD 59213) and 2021, January 26 (MJD 59240); \textit{Middle panel}: \textit{Swift}/UVOT observations of the 2021 flare. Upper limits are not plotted, for clarity. All UVOT magnitudes are AB magnitudes. With solid, dashed, dotted and dashed-dotted lines, the quiescent levels in uvw1, uvw2, uvm2 and u (respectively) of the source are drawn. The quiescent levels have been estimated by averaging UVOT archival data, starting January 2012. \textit{Bottom} panel: \textit{Swift}/XRT and {\it NICER} observations of the 2021 flare.}
    \label{fig:flare_lc}
\end{figure}

In the end, a total of 109, 292, 36, 163 and 294 reliable magnitudes in the $g'$, $V$, $r'$, $R$ and $i'$ bands (Tab. \ref{tab:wavelengths}), respectively, have been obtained during our long-term optical monitoring of Cen X-4 with LCO (Fig. \ref{fig:long_monitoring}).

\subsection{Optical and near infrared observations with REM}
Cen X-4 was observed on January 5, 2021 (MJD 59219) with the 60cm Rapid Eye Mount (REM; \citealt{Zerbi2001}; \citealt{Covino2004}) telescope (La Silla, Chile). Strictly simultaneous, 300s integration time observations have been obtained using the optical SDSS $g'r'i'z'$ filters (Tab. \ref{tab:wavelengths}), for a total of 9 observations per filter. Images were reduced using standard procedures (bias subtraction and flat-field correction), and aperture photometry was performed on the stars in the field using {\tt PHOT} in {\tt IRAF}\footnote{IRAF is distributed by the National Optical Astronomy Observatory, which
is operated by the Association of Universities for Research in Astronomy, Inc.,
under cooperative agreement with the National Science Foundation.}. Photometry was then flux-calibrated using APASS\footnote{\url{http://www.aavso.org/download-apass-data}} stars in the field \citep[][]{Henden2019}.

The system was then observed again on May 22, 2021 (MJD 59356) with REM. Observations were acquired in the optical SDSS $g'$, $r'$, $i'$, $z'$ bands, strictly simultaneously (90s integration, for a total of 26 images/filter). Reduction and analysis of the optical data was performed as described above.
At the same time, NIR (2MASS $JHK$ bands) observations were acquired with the REMIR camera mounted on REM, alternating the filters, performing 15s integration exposures. A total of 90 images/filter were acquired. Dithering of the images was performed with the aim of evaluating the variable contribution of the sky, which was then subtracted from each image. Images were then combined 5 by 5 to increase the signal to noise. Flux calibration of the NIR images was performed against a group of 2MASS stars in the field.


\subsection{Swift X-ray and optical/UV monitoring}

The Neil Gehrels Swift Observatory \citep[hereafter \emph{Swift};][]{buhino2005} observed Cen X--4 16 times between 2020 December 28 and 2021 January 23 with the X-Ray Telescope (XRT; \citealt{Burrows2005}) and Ultraviolet and Optical Telescope (UVOT; \citealt{Roming2005}) instruments. For the XRT we only analyzed data obtained when the instrument was in Photon Counting mode; as the source was too faint to be detected in the short Window Timing mode exposures. For each XRT observation, we extracted source spectra from a circular aperture with a radius of 20 pixels ($\sim$47\arcsec) centred on the source. Background-only spectra were extracted from an annulus with inner and outer radii of 40 and 60 pixels ($\sim$94\arcsec\ and $\sim$141\arcsec), respectively, also centred on the source. From the background-subtracted spectra, we created a 0.5--10 keV light curve (with one data point per observation), correcting for changes in the effective area between observations that resulted from differences in how bad columns affected the source counts.


The UVOT instrument observed Cen X-4 during the 2020/2021 flare, using all available filters ($v$, $b$, $u$, $uvw1$, $uvm2$, $uvw2$), for a total of 14 epochs between MJD 59211 (2020 December 28) and MJD 59237 (2021 January 23). The data were analysed using the {\tt uvotsource} HEASOFT routine, defining as the extraction region a circular aperture centred on the source with a radius of 3 arcsec, and as
background a circular aperture (away from the source) with radius of 10 arcsec.
Several detections have been obtained, in addition to some upper-limits in all bands. The light curves are shown in the mid-panel of Fig. \ref{fig:flare_lc}.

\subsection{{\it NICER} monitoring}

The Neutron Star Interior Composition Explorer ({\it NICER}; \citealt{Gendreau2012}; \citealt{gearad2016}) observed Cen X-4 extensively in early 2021. We analyzed all observations made between January 1 and February 19. The observations, each comprised of one or more good-time-intervals, were reprocessed using the {\tt nicerl2} script that is part of the {\tt NICERDAS} package in {\tt HEASOFT v6.28}, using calibration version 20200722. 

Spectra were extracted for each good-time-interval (GTI) using the tool {\tt nibackgen3C50}, which also creates background spectra \citep{remillard2021}. For some GTIs, the parameters used to calculate background spectra could not be matched with the pre-calculated library of background spectra that is used by {\tt nibackgen3C50}. In those cases the GTI was excluded from our analysis, leaving a total of 186 spectra, with exposure times ranging from 51 s to 2627 s. 

Background-subtracted light curves in the 0.5-10 keV band were extracted from the spectra, with each data point representing the average count rate of a single GTI. Inspection of the resulting light curve revealed strong flaring during the time interval of MJD 59240--59250 (January 26 to February 5, 2021), that were likely due to residual background. By filtering out GTIs for which the background count rate in the 0.5-10 keV band was $>$0.5 counts\,s$^{-1}$ these ``flaring'' episodes were almost completely removed. Several suspicious outliers on MJD 59241-59242 and after MJD 59260 were removed manually.


\section{Results}\label{Results_Section}
\subsection{The long-term optical monitoring}


The long-term optical light curves obtained with LCO are shown in Fig. \ref{fig:long_monitoring}. Strong variability is observed, characterized by the emission of optical flares or dips of the order of up to $\sim 0.5$ mags on timescales of 1-2 months. A similar level of activity was previously reported in the optical by \citet{Zurita2003} and in the X-rays/UV by \citealt{Campana2004} and \citet{Bernardini2013}. Moreover, a decreasing trend in the average optical flux is observed in the long term monitoring, up to $\sim$ MJD 58016 (September 20, 2017). After that date, the average flux gradually increased until right before the start of the 2020 period of Sun constraint (i.e. September 2020).
Any long term variability in the optical light curve of LMXBs is typically related to an evolution of the accretion disk (see e.g. the case of V404 Cyg; \citealt{Bernardini2016_precursor}), or, rarely, of the jet (as in the case of Swift J1357.2-0933; \citealt{Russell2018}; Caruso et al. in preparation). The companion star contribution is instead expected to exhibit a double-humped ellipsoidal modulation at the orbital period of the source \citep{Orosz1997}.
We know from previous studies that jets are unlikely to contribute to the quiescent optical emission of Cen X-4 \citep{Baglio2014}, and therefore we will principally focus on the accretion disc emission.
To estimate the level of variability of the stable accretion disc, we first determined a flux threshold for the emission of flares. Flares likely originate in the accretion disc, and are probably due to variability in the accretion rate, which happens on the viscous time scale (days--weeks), or could be related to irradiation and have timescales of seconds-minutes. 

\begin{figure}
    \centering
    \includegraphics[scale=0.4]{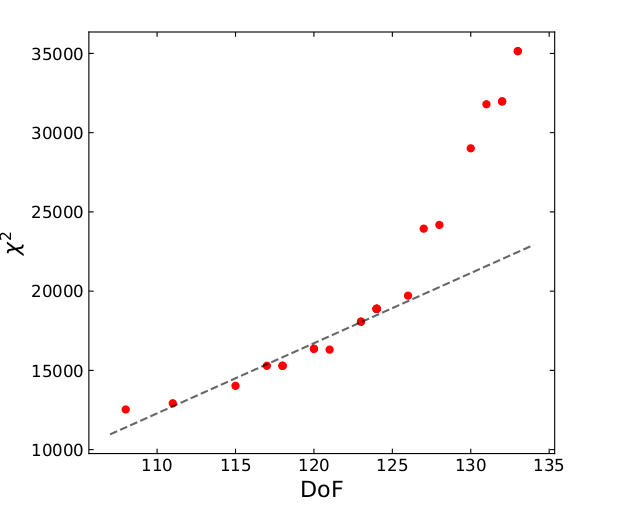}
    \caption{$\chi^2$ of the fit of a double sinusoidal function plus a constant to the $R$ band light curve of Cen X-4 against the number of dof in the fit for the different magnitude thresholds that we considered. Superimposed, we plotted a dashed line corresponding to the linear fit of the lowest dof points, before the transition to a steeper correlation happens. }
    \label{fig:chi2}
\end{figure}

Following the work by \citet{Jonker2008}, performed on the accreting millisecond X-ray pulsar IGR J00291+5934, we first folded the light curves on the known orbital period of the source (0.6290630 days; \citealt{McClintock1990}); then we established a possible magnitude threshold brighter than which the points are assumed to be flares from the disc, and we removed all the magnitudes that were brighter than the threshold in each band. We further binned these points in 20 bins of orbital phase of equal width. To better approximate the double-humped ellipsoidal modulation of the companion star emission, we performed a non-linear weighted least-squares fit to the binned magnitudes ($m$) vs. phase ($x$) data with a double sinusoidal function plus a constant: m=$C+ A_1 \, \sin(2\pi\,(x-\Phi)/0.5-\pi/2)+A_2\, \sin(2\pi\,(x-\Phi)-\pi/2)$, where $C$ is a constant magnitude, $\Phi$ is the phase corresponding to the inferior conjunction of the companion star, and $A_1$ and $A_2$ are the semi-amplitudes of the two oscillations; we note that one oscillation has a fixed double periodicity with respect to the other, and the free parameters of the fit are $C$, $A_1$, $A_2$, $\Phi$. We computed the $\chi^2$ and the degrees of freedom (dof) of the fit. We then changed the threshold value and repeated the above steps. Eventually, we plotted our results in a $\chi^2$ against number of dof plot, for all bands (see Fig. \ref{fig:chi2} for the $R$ band plot); as in \citet{Jonker2008}, the relation is linear until a certain level, then it deviates from the linear correlation. We therefore took the point where the deviation occurs as the threshold level for the flares: $V=17.98$ mag, $R=17.27$ mag and $i'=17.15$ mag. \\

\begin{figure}
    \centering
   \includegraphics[width=8cm]{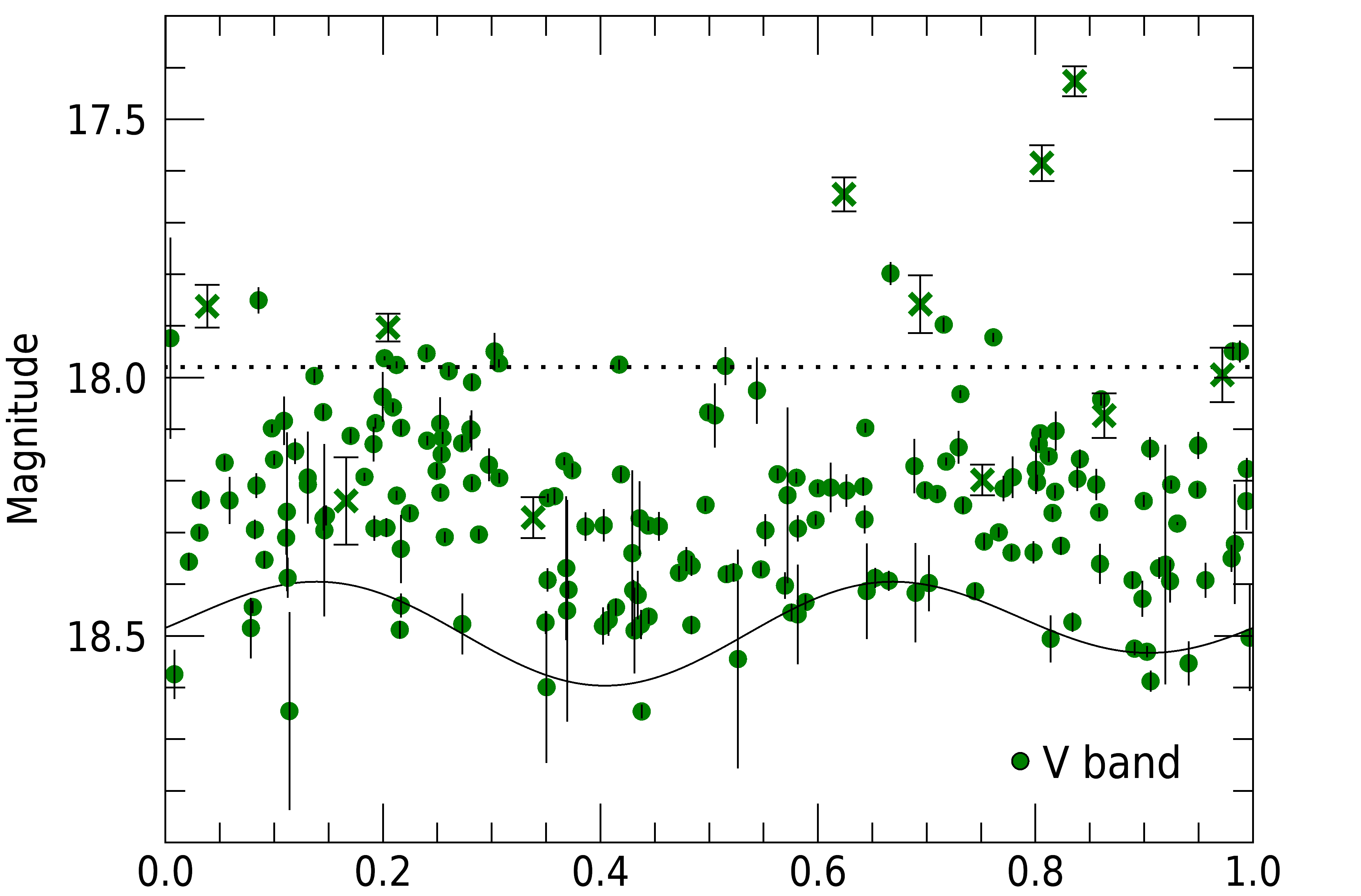}
    \includegraphics[width=8cm]{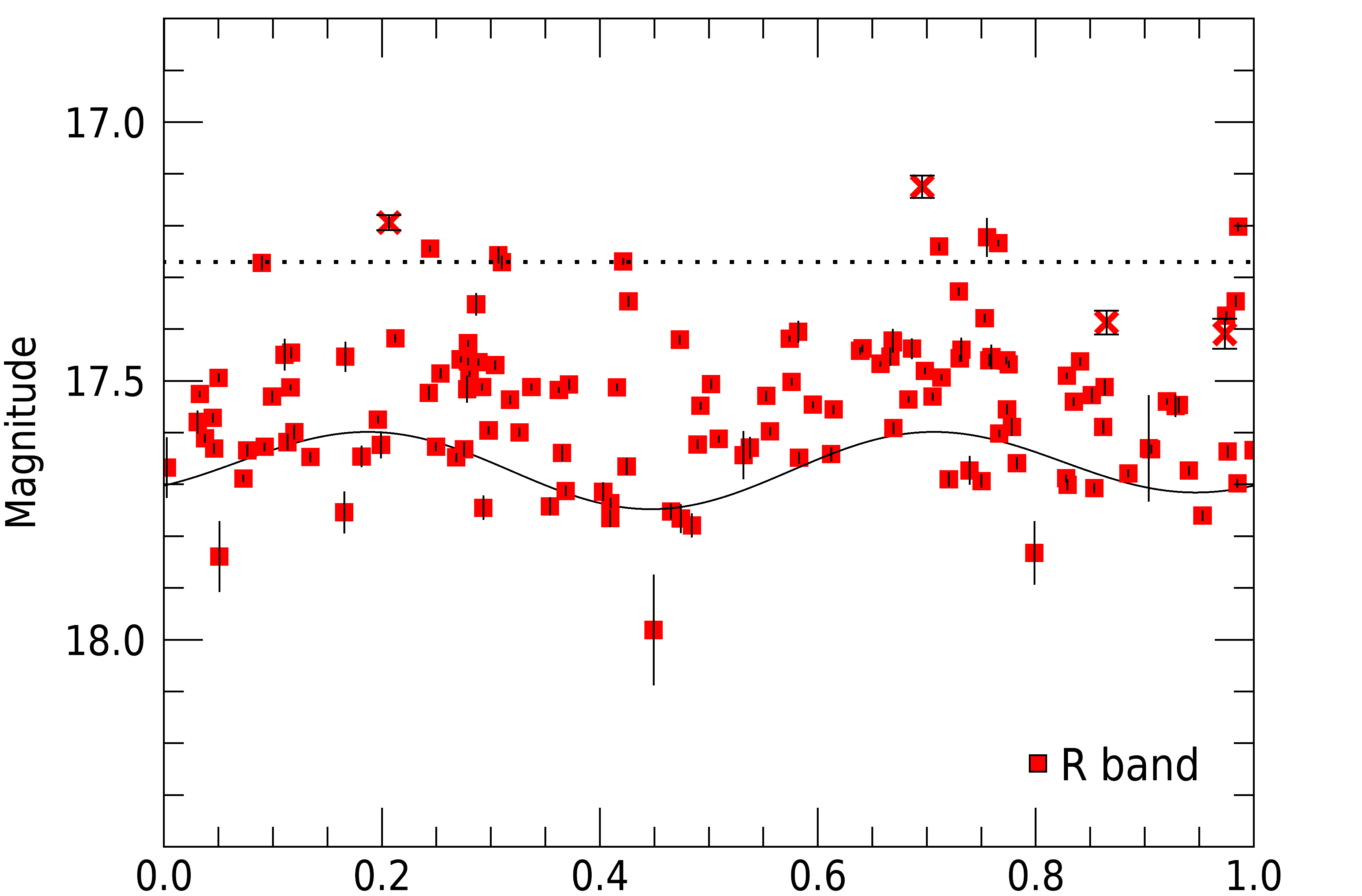}
    \includegraphics[width=8cm]{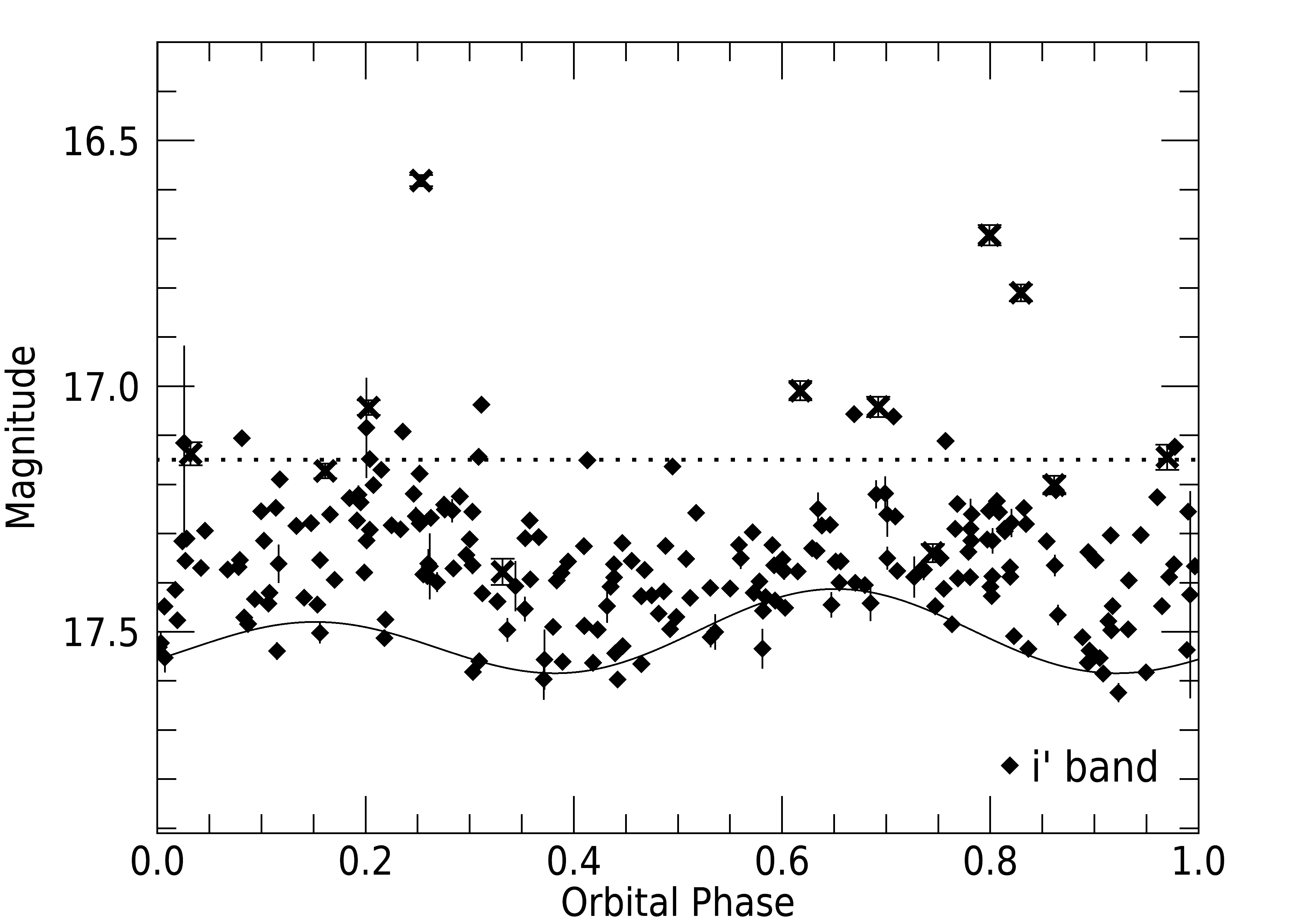}
    \caption{From top to bottom, $V$, $R$ and $i'$-band light curves folded on the $\sim 15.1$ hr orbital period \citep{McClintock1990}. Orbital phases are calculated according to the ephemeris by \citet{McClintock1990}. The black solid curve shows the best fit to the lower envelope in each case, considering a simple model for the expected ellipsoidal modulation. 
    In each panel, an horizontal black dashed line indicates the magnitude threshold for the flares (see text). All points lying above the dashed line have therefore to be considered as flares. All the observations during the 2020/2021 misfired outburst are plotted with an 'x' symbol for comparison. }
    \label{fig:orbital_modulation}
\end{figure}

Once all flares were excluded, the folded light curves show a modulation, which is expected from the companion star, and some scatter (the errors from the photometry are much smaller than the observed scatter; Fig. \ref{fig:orbital_modulation}). In all filters, the semi-amplitude of the modulations is low ($\sim0.1$ mag).

We first performed a fit with the double sinusoidal function model, in order to evaluate the parameters of the modulation. However, the light curves still have a significant contribution coming from the accretion disc. In order to isolate the modulation from the companion star, we estimated the lower envelope of the modulation following \citet{Pavlenko1996} and \citet{Zurita2004}. We divided the $V, R, i'$ light curves into 10 identical phase bins; for each bin, we found the minimum brightness; we defined the lower envelope emission as all the observations that differ from this minimum by at most twice the average uncertainty of the 10 faintest observations in the bin. We then performed the fit of the lower envelope of the modulation with the double sinusoidal model, fixing the parameters of the modulation to those obtained for the whole light curves, after the flares removal (solid line in Fig. \ref{fig:orbital_modulation}). The constant magnitude of the modulation corresponds to $V=18.48 \pm 0.01$, $R=17.66 \pm 0.01$, $i'=17.51 \pm 0.01$.

The lower envelope of the modulation is plotted as a solid line in Fig. \ref{fig:orbital_modulation}.

\begin{figure}
    \centering
    \includegraphics[scale=0.5]{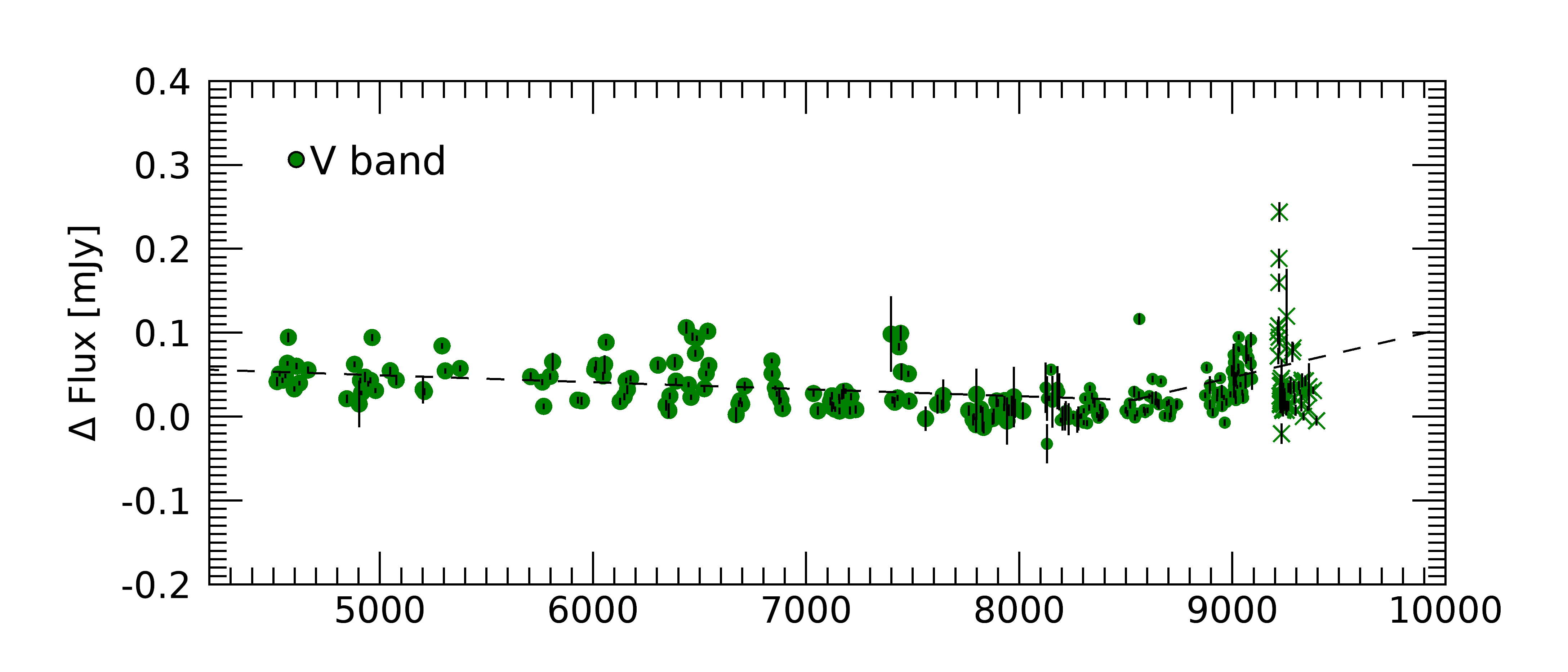}
    \includegraphics[scale=0.5]{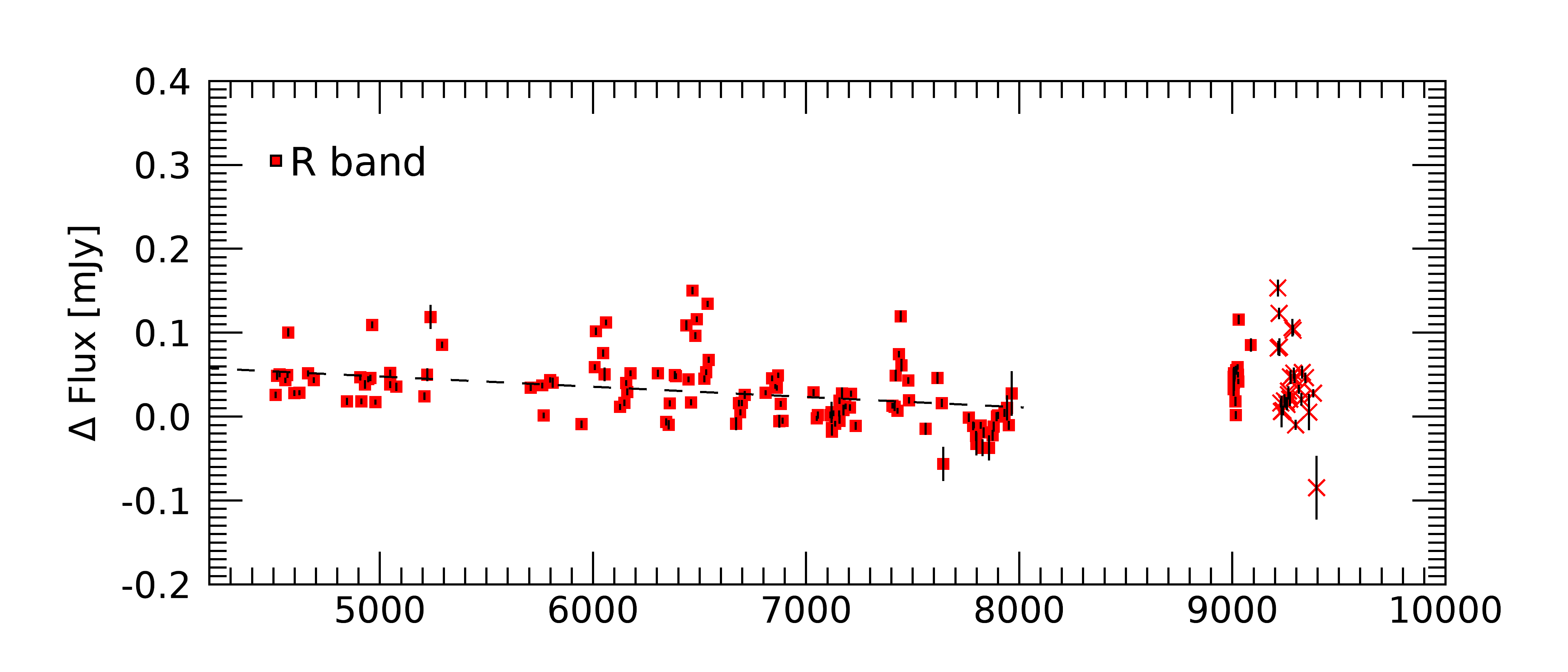}
    \includegraphics[scale=0.5]{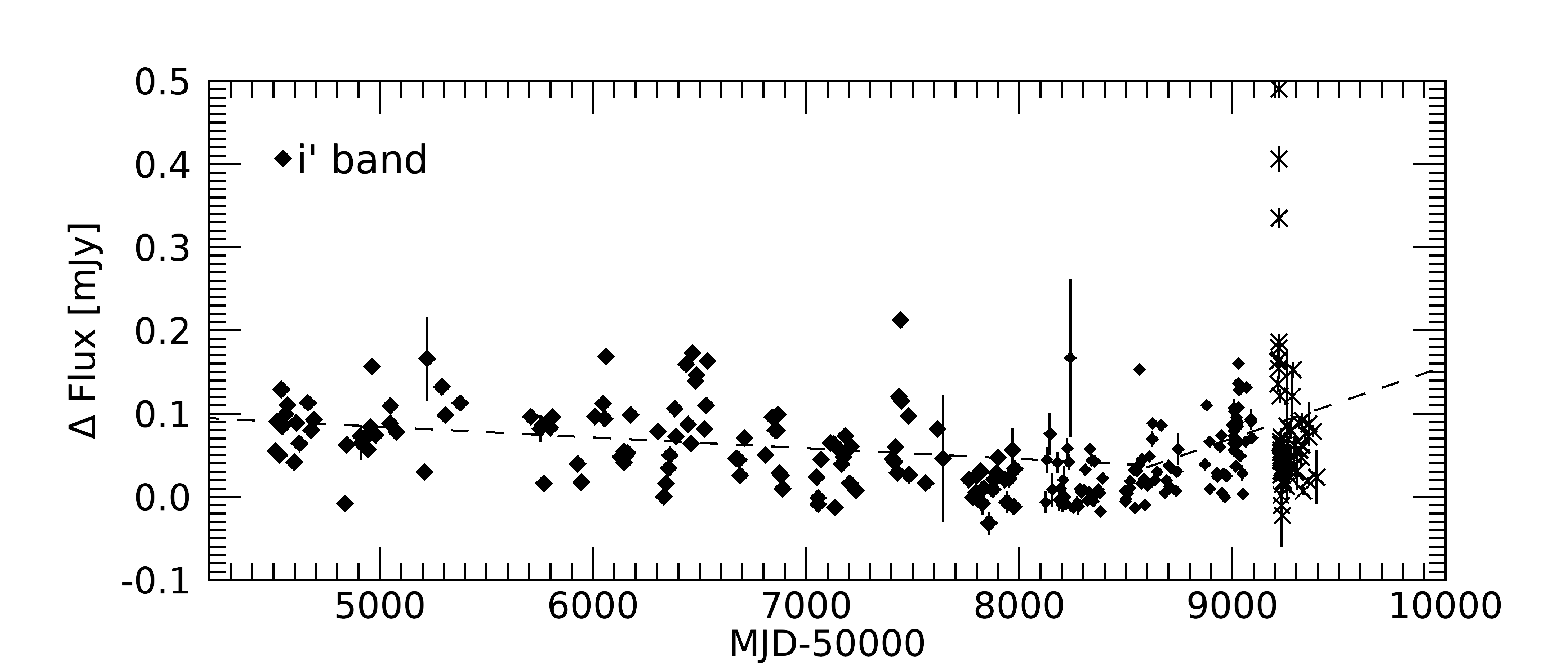}
   \caption{From top to bottom, $V$, $R$ and $i'$-band light curves of the residuals, obtained after the subtraction of the sinusoidal modulation from the original light curves. Only the points from before the beginning of the 2020 Sun constraint are shown. Superimposed with dashed lines, the linear fits of the long-term trends are shown, where possible. The observations acquired during the 2020/2021 misfired outburst are plotted with an 'x' symbol for comparison.}
    \label{fig:residuals}
\end{figure}

We then subtracted the contribution of the lower envelope (constant+modulation) from every data point, with the aim of isolating the emission from the accretion disc. 
We converted the resulting magnitudes into flux densities (mJy), and built a light curve of the non-stellar flux densities during the last 13 years of observations. The result is presented in Fig. \ref{fig:residuals}, and clearly shows a downward trend of the flux emitted from the disc before $\sim$ September 20, 2017 (MJD 58016), followed by an upward trend after this date. We performed a weighted least squares fit with a constant plus a linear function ($C+A\,t$, where $C$ is a constant flux, $t$ is time expressed in MJD, and $A$ is the gradient of the line) of the two trends separately for each band (no upward trend fitting was possible for the $R$-band, due to the lack of data after MJD 58016). We note that the inclusion of the linear function improves the fit in all cases with a $> 10 \sigma$ significance, according to an $F$-test.

The results of the fit show a decrease of flux of $(0.83\pm 0.02)\times 10^{-5}$ mJy/day, $(1.22\pm 0.02)\times 10^{-5}$ mJy/day and $(1.28\pm 0.03)\times 10^{-5}$ mJy/day in $V$, $R$ and $i'$ band, respectively, before MJD 58016, and an increase of flux of $(5.85\pm0.08) \times 10^{-5}$ and $(8.67\pm0.11) \times 10^{-5}$mJy/day in $V$ and $i'$ band, respectively, after MJD 58016. 
The upward trend is therefore $\sim7$ times steeper with respect to the downward one, both in $i'$ and $V$ band.

\subsection{The 2020/2021 flare}


After the Sun constraint ended, on 2020 December 30 (MJD 59213), Cen X-4 was found to be significantly brighter at optical wavelengths than before \citep{Saikia2021}, with a brightening of $0.57\pm0.12$ and $0.42\pm0.09$ mag in $V$ and $i'$ band, respectively, compared to the previous point. 
During the first days of activity, the rise in the optical emission was found to be steep, with a flux increase of $\sim0.3$ mags and $\sim0.8$ mags in $\sim6$ days (i.e. until 2020 January 5; MJD 59219) in $V$ and $i'$ band, respectively.
From our long-term monitoring of Cen X-4 with LCO, the modulation of the source has a $\sim 0.1$ mag semi-amplitude, which is much smaller than what is required to explain the amplitude of the variability. 
However, instead of undergoing a full outburst, the flare peaked on MJD $\sim 59219-20$ (2021 January 5-6) in all optical bands, and then started to fade rapidly, at a similar, steep rate as during the rise (losing $\sim 1-1.2$ mag in $\sim 8-9$ days in all bands), and reached quiescent levels again on MJD $\sim 59228$ (2021 January 14), $\sim 8$ days after the peak. Since a proper outburst did not have the chance to start, we classify this peculiar activity as ``misfired outburst''. 

Looking at Fig. \ref{fig:residuals}, we note that soon before the beginning of the Sun constraint, a few detections were lying above the quiescent level indicated by the linear fit at all wavelengths. However, this flux increase looks comparable to the amount of activity typically observed during quiescence for Cen X-4 (see Fig. \ref{fig:residuals}). We consider therefore unlikely that these points are marking the beginning of the misfired outburst, which likely started during the Sun constraint, or at the end of it. 

\subsubsection{Short term optical variability}
\begin{figure*}
    \centering
    \includegraphics[width=9cm, angle=0]{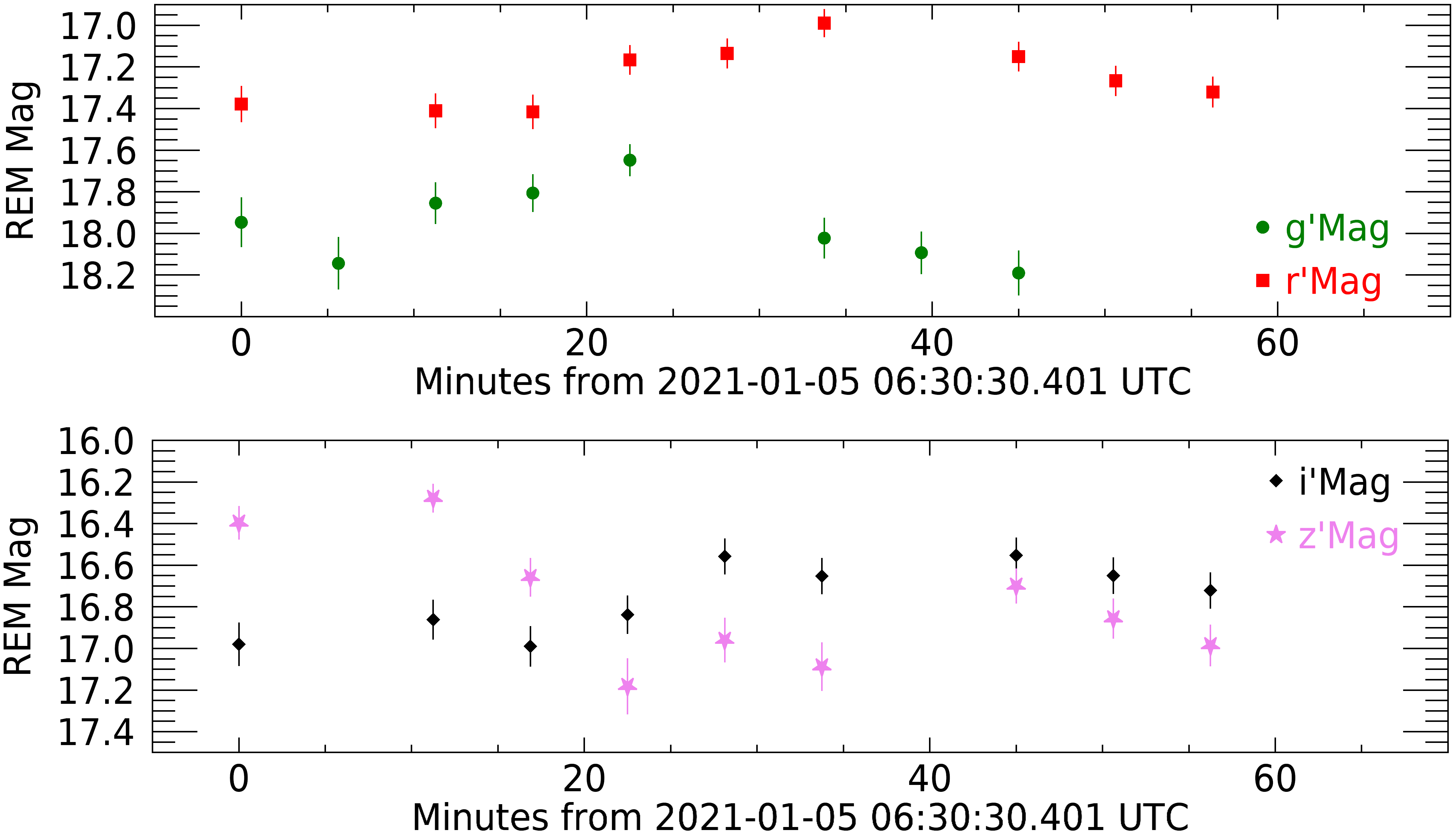}
    \includegraphics[width=8.9cm, angle=0]{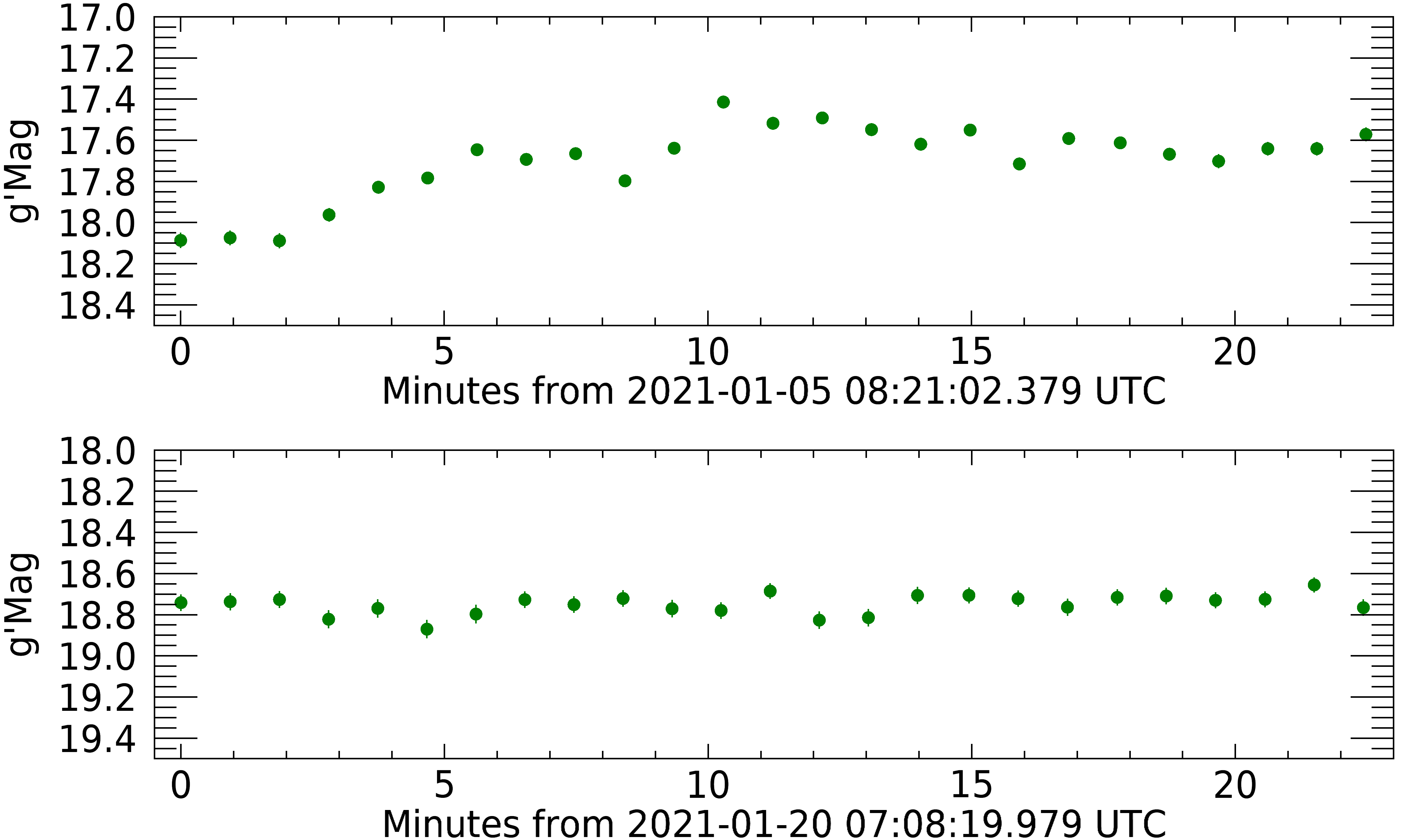}
    \caption{\textit{Left}: $g'$, $r'$ (top panel), $i'$, $z'$ (bottom panel) light curves obtained with REM on 2021, Jan 05 (MJD 59219), showing minute-timescale variability. \textit{Right}: $g'$-band light curves obtained with LCO on 2021, Jan 05 and 20 (MJD 59219 and 59234, respectively), showing minute-timescale variability.}
    \label{fig:short_ts_lc}
\end{figure*}

The REM observations performed during the misfired outburst on 2021 January 5 (MJD 59219) resulted in optical light curves showing variability (Fig. \ref{fig:short_ts_lc}, left panel). 
Following \citet{Vaughan2003}, we evaluated the fractional root-mean-square (rms) of the light curves in order to quantify the variability in the $i'$, $r'$, $g'$ bands, and we measured a fractional rms of $(12.9\pm2.9)\%$, $(12.1\pm2.3)\%$, $(15.1\pm 3.4)\%$ in $i'$, $r'$, $g'$-band, respectively. The intrinsic variability is therefore comparable in all bands.
In $z'$-band, a dramatic variability is observed, with $\sim 1$ mag difference between the lowest and highest point of the light curve (and a fractional rms of $28.6\pm2.8\%$. However, very  similar variability is also observed for a comparison star of similar brightness in $z'$-band, so we tend to attribute it to the strong fringing of the $z'$-band images. 

A similar, higher significance variability is observed in the 22.5 min duration light curve obtained on the same day (MJD 59219; Jan 5) with LCO in the $g'$-band, which has a fractional rms of $(15.5\pm0.5)\%$, for a light curve with a time resolution of $\sim 56$s. A $g'$-band light curve with the same time-resolution, obtained at the end of the flaring episode with LCO on MJD 59234 (2021 January 20), has a significantly lower short timescale variability (fractional rms of $(2.8\pm0.9)\%$; Fig. \ref{fig:short_ts_lc}, right panel). 
Despite the value of the fractional rms being comparable, this variability is observed on much longer timescales (minutes; Fig. \ref{fig:short_ts_lc}) with respect to sources like the BH XRBs GX 339-4 or MAXI J1535-571 (seconds, or less), for which the variability was attributed to the presence of a flickering jet (e.g. \citealt{Gandhi2010}; \citealt{Baglio2018}). It is therefore unlikely that the observed optical variability can be attributed to the emission of jets in the system.

\subsubsection{X-rays}\label{X-ray_sec}

The X-ray coverage of the 2020/2021 flare started on MJD 59211 (December 28 2020). The {\it Swift/XRT} and {\it NICER} light curves in Figure \ref{fig:flare_lc} show that the X-ray flare peaked around the same time as the optical, between MJD 58918 and MJD 59221 (Jan 4 and Jan 7 2021). The {\it NICER} light curve (which has the higher time resolution) reveals strong variability (by factors of 2--3) on a time scale of hours near the peak of the flare. Analysis of archival data shows that the X-ray peak count rates observed during the 2020/2021 flare were a factor $\sim$2 ({\it Swift/XRT}) and $\sim$2.5 ({\it NICER}) times higher than the maximum count rates of Cen X-4 in observations made prior to December 2020, when the source was in quiescence.

Spectra obtained from most of the {\it Swift} observations and {\it NICER} GTIs were not of sufficient quality to perform detailed spectral fits.  Using XSPEC V12.11.1 \citep{ar1996}, we performed a fit to the {\it NICER} spectrum with the highest count rate (MJD 59220.373038, the first GTI of observation 3652010501, with an exposure time $\sim$1250 s). The main goal of this spectral fit was to obtain a reliable count rate to flux conversion factor that can be used to estimate the outburst flux (see Sec. \ref{sec:dim}), under the assumption that the spectral shape didn't change significantly during the outburst. The {\it NICER} spectrum was rebinned to a minimum of 30 counts per spectral bin so that $\chi^2$ fitting could be employed. Following \citet{cackett2010} and \citet{chakrabarty2014}, who studied the variable quiescent spectra of Cen X-4, we fit the 0.5--10 keV spectrum with a continuum model comprised of a thermal and a non-thermal component. For the thermal component we used the neutron-star atmosphere model of  \citet{heinke2006} ({\tt nsatmos} in XSPEC) and for the non-thermal component we used a power-law; the band pass of {\it NICER} did not extend high enough to test more sophisticated models for the non-thermal component (as was done in \citealt{chakrabarty2014}, for example). Interstellar absorption was modelled with the {\tt tbabs} model in XSPEC, with  the abundances set to {\tt WILM} and cross sections to {\tt VERN}. For the {\tt nsatmos} component we fixed the neutron-star mass to 1.9 $M_\odot$ \citep{shwadh2014} and the  distance to 1.2 kpc. The model fits well ($\chi^2$=145 for 145 degrees of freedom); we obtain an $n_{\rm H}$ of 6.52(1)$\times10^{20}$ cm$^{-2}$, a neutron-star temperature log($T_{\rm nsa}; K)$=6.24$\pm$0.05, a neutron-star radius of 9.6$\pm$1.3 km, and a power-law index of 0.73$\pm$0.18. The unabsorbed 0.5--10 keV flux was (1.95$\pm$0.10)$\times10^{-11}$ erg\,cm$^{-2}$\,s$^{-1}$ (corresponding to a luminosity of $\sim3.4\times10^{33}$ erg\,s$^{-1}$ at 1.2 kpc), with the power-law contributing $\sim$50\% in the 0.5--10 keV band. This gives count rate to flux conversion factor of $\sim$2.6$\times10^{-12}$ erg\,cm$^{-2}$\,cts$^{-1}$. The power-law index of 0.73 is very low compared to  NS LMXBs in a slightly higher luminosity range \citep[$>10^{34} \rm erg\,s^{-1}$; see, e.g.,][]{wijnands2015,stoop2021} where the index is around 2.5, but it is consistent with the lowest values found by \citep{cackett2010} for Cen X-4. We note that a fit with a single power-law does not perform well, yielding  $\chi^2$=246 for 147 degrees of freedom  (power-law index of 3.37$\pm$0.07 and  $n_{\rm H}$ of 2.6(2)$\times10^{20}$ cm$^{-2}$). 

\section{Discussion}

\subsection{Long-term optical monitoring}
We have been monitoring the long-term quiescent optical behaviour of Cen X-4 for almost 13.5 years, since Feb. 14th 2008. After taking into account the modulation due to the companion star, we isolated the accretion activity of the source and observed a linear downward trend followed by a steeper upward trend during quiescence. From the gradual optical brightening detected in the long term lightcurve of Cen X-4, \citet{Waterval2020} predicted that Cen X-4 might enter an outburst in the near future. Subsequently, flaring activity of the source was detected both at optical (\citealt{Saikia2021}, \citealt{Baglio2021}) and X-ray wavelengths \citep{Eijnden2021_1}.

The DIM predicts a continuously increasing optical flux during quiescence \citep[][]{Lasota2001}, but observations of both LMXBs and dwarf novae typically show a constant or decreasing flux with time, as we detect for Cen X-4 in our optical monitoring. A very similar behaviour was reported for the BH XRB V404 Cyg \citep[][]{Bernardini2016_precursor}, where a 0.1 mag decrease in brightness over $\sim 2000$ days was observed, and linked to changes in the accretion rate from year to year (as is likely the case for Cen X-4, too).
This decay was then followed by a low-amplitude, relatively fast enhancement of optical emission (0.1 mag increase over $\sim 1000$ days), that was an indication of an increase in the mass accretion rate, which eventually culminated in the 2015 outburst of the source. Other X-ray transient sources where a slow and significant optical rise has been seen together together with an outburst, are the BH XRBs GS 1354-64 (BW Cir; \citealt{Koljonen2016}) and Swift J1357.2-0933 \citep{Russell2018}. Similarly, a slow optical rise during quiescence was observed for the BH XRBs H1705-250 and GRS 1124-68 (see \citealt{Yang2012} and \citealt{Wu2016}, respectively; see also Tab. 1 of \citealt{Russell2018} for a summary), although no new outburst has yet been detected for these sources.

Even though on different timescales, an optical precursor to an outburst has recently been observed also for the NS LMXB SAX J1808.4-3658 \citep{Goodwin2020}, that underwent a complete outburst in August 2019. The optical magnitude was observed to fluctuate by $\sim 1$ magnitude for $\sim$ 8 days before the proper outburst rise was initiated in the optical. This optical precursor can have several possible origins: an enhanced mass transfer from the companion star, which would then help triggering the outburst; instabilities in the outer disc, which could lead to heating fronts propagating through the entire disc, that would contribute to igniting the outburst; changes in the pulsar radiation pressure, the compact object being a millisecond pulsar. 
Similarly, signatures of enhanced optical activity soon before the onset of an outburst have been suggested also for the NS LMXB IGR J00291+5934, whose optical lightcurve is dominated by flaring and flickering activity prior to the start of an outburst, completely hiding the sinusoidal modulation of the companion star \citep{Baglio2017}.

\begin{figure}
    \centering
\includegraphics[width=9.5cm]{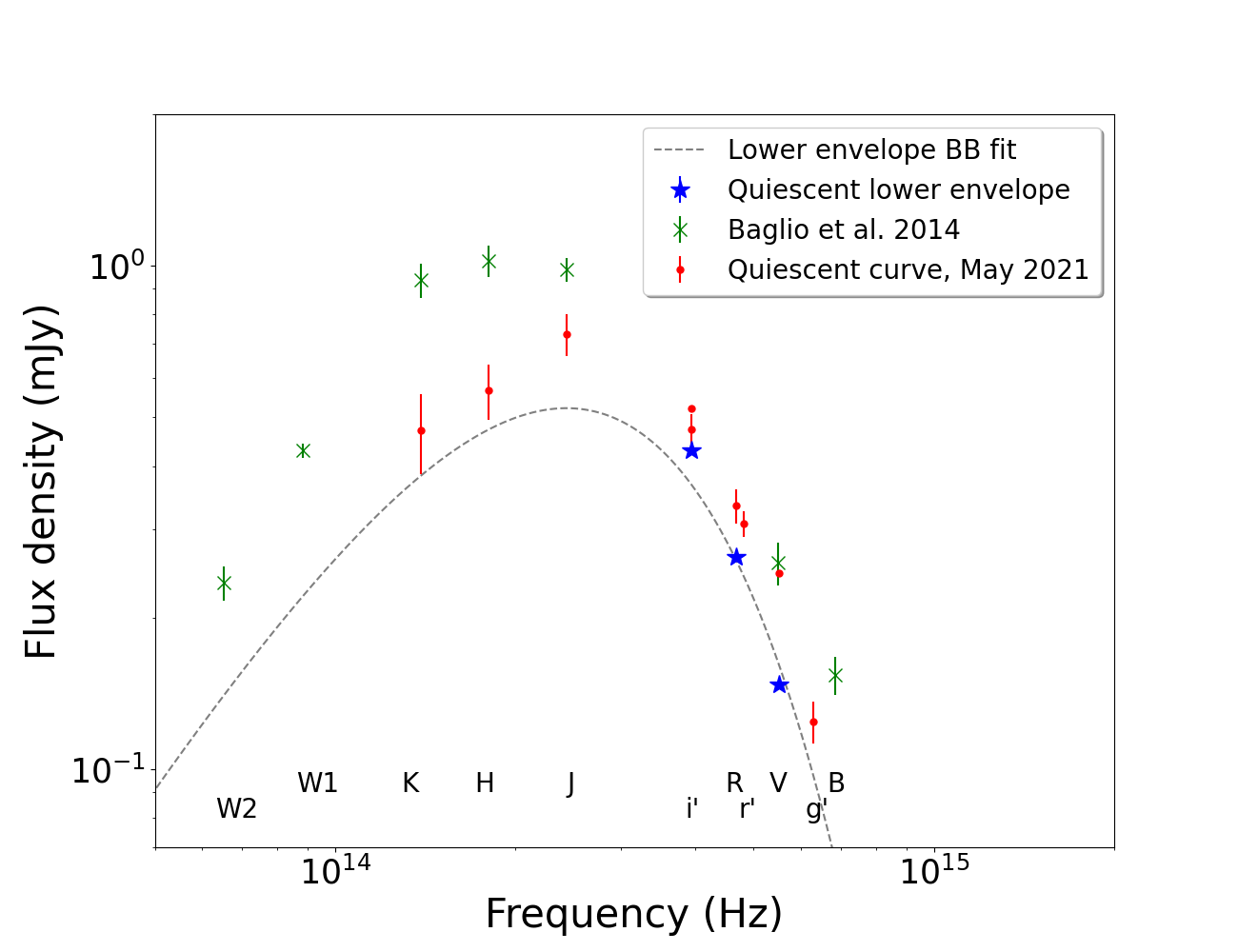}
\includegraphics[width=9.5cm]{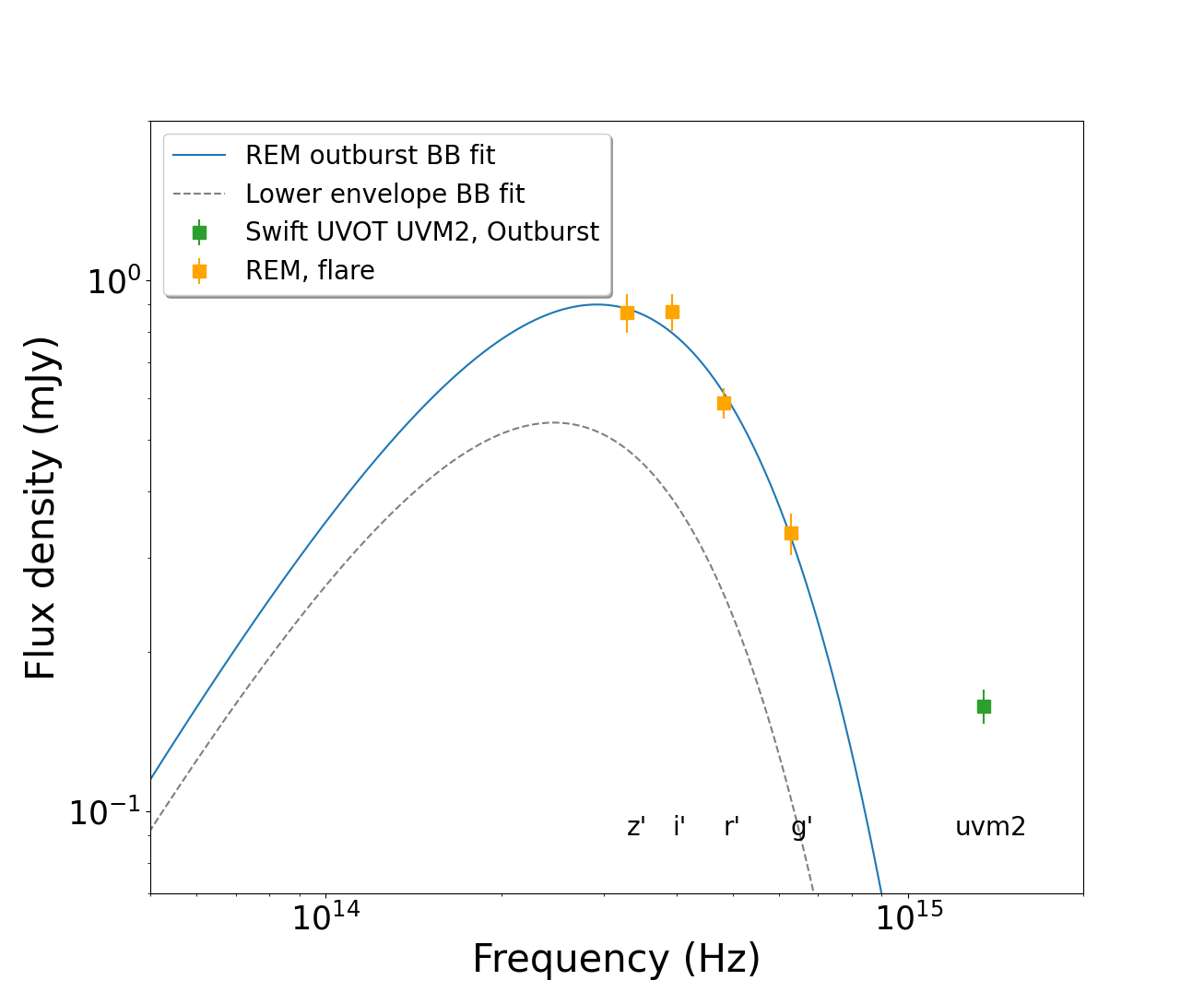}
    \caption{\textit{Top}: Quiescent de-reddened SED of Cen X-4 obtained in May 2021 using REM and LCO data (red dots). With green `x', the quiescent curve published in \citet{Baglio2014} is plotted, for comparison. Blue stars instead represent the fluxes obtained as the average of the lower envelope emission in the long-term monitoring of Cen X-4 during quiescence (Fig. \ref{fig:orbital_modulation}). Superimposed, the fit of the lower envelope emission with a non-irradiated blackbody. \textit{Bottom}: Average de-reddened SED of Cen X-4 during the recent misfired outburst, based on REM strictly simultaneous observations acquired on January 5, 2021 (orange squares) and \textit{Swift}/UVOT (same date, uvm2 filter; green square). Superimposed, the fit of the REM points with an irradiated star black body (blue, solid line). With a dashed grey line, the blackbody fit of the quiescent lower envelope is plotted (from the top panel), for comparison purposes.}
    \label{fig:comparison_SED}
\end{figure}

\begin{figure}
    \centering
    \includegraphics[width=9cm]{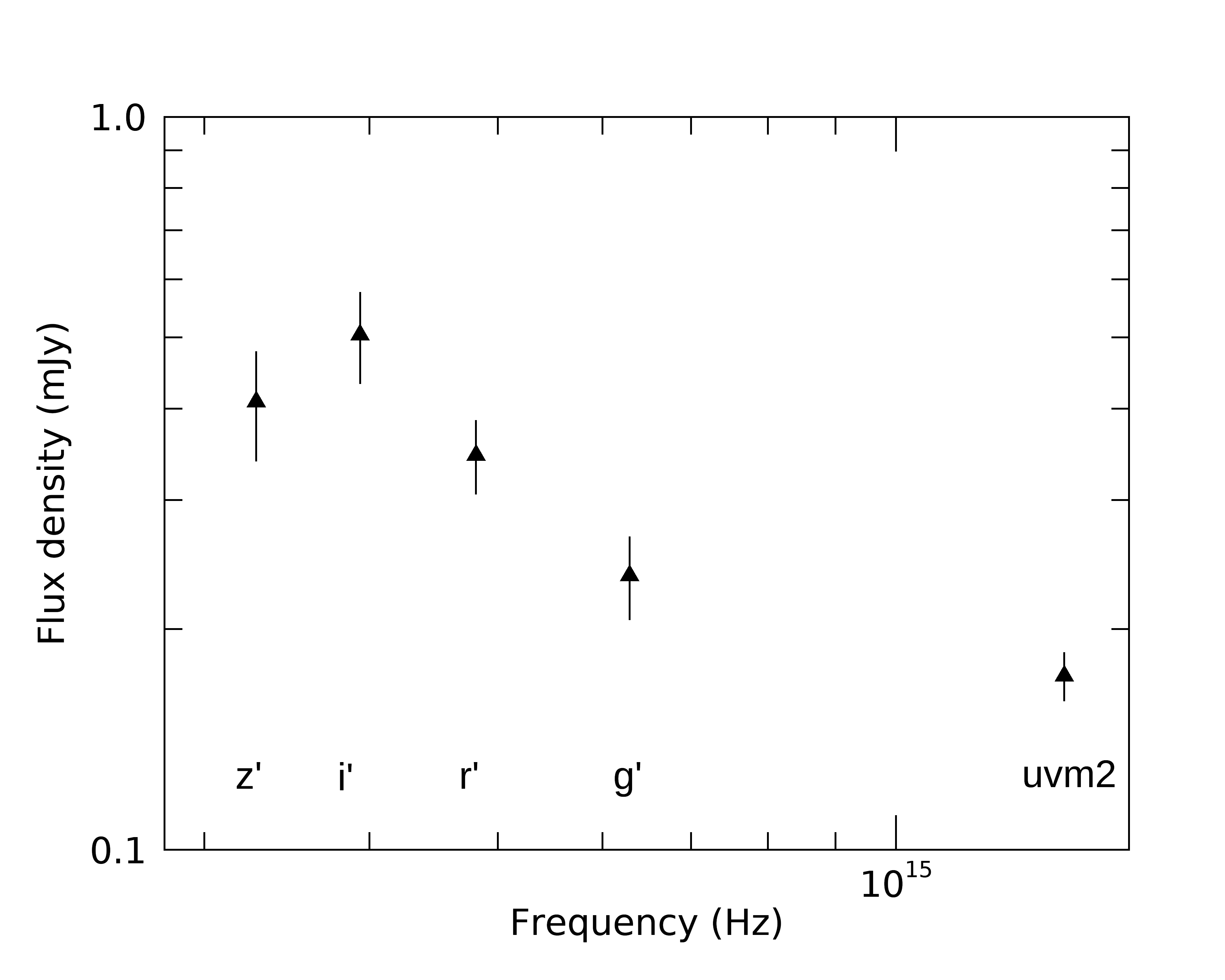}
    \caption{Residual fluxes of Cen X-4 during the misfired outburst after the subtraction of the companion star emission, obtained as the black body fit to the lower envelope emission (see Fig. \ref{fig:comparison_SED}, upper panel).}
    \label{fig:subtracted_SED}
\end{figure}

The optical flux enhancement leading towards the flaring activity observed for Cen X-4 supports the DIM with irradiation and disc evaporation/condensation \citep{Dubus2001}, which explains the evolution outburst-quiescence mechanisms at all wavelengths in an X-ray binary. The DIM predicts that during quiescence, the cold disc accumulates mass from the companion star via Roche lobe overflow, and that causes the gradual brightening of the disc in optical wavelengths \citep{Lasota2001}. Generally, an outburst is expected to occur when the accretion disc reaches a critical density. The disc temperature then increases, causing hydrogen in the disc to ionize. This heating front is propagated through the disc closer to the inner accretion flow, causing enhancement of activity in higher-energy wavebands like X-rays, and the outburst starts.

The gradual brightening of Cen X-4 in quiescence can therefore be explained with matter slowly accumulating in the accretion disc and getting optically brighter. The amount of matter in the disc, increasing year after year, could account for the increasing optical flux that is observed (similarly to what happened for V404 Cyg; \citealt{Bernardini2016_precursor}). However, for Cen X-4 the optical and X-ray flaring activity did not lead to the ignition of a proper outburst, for reasons that we will discuss in the next sections. 


\subsection{The misfired outburst}

\begin{figure*}
\centering
\includegraphics[width=9.1cm]{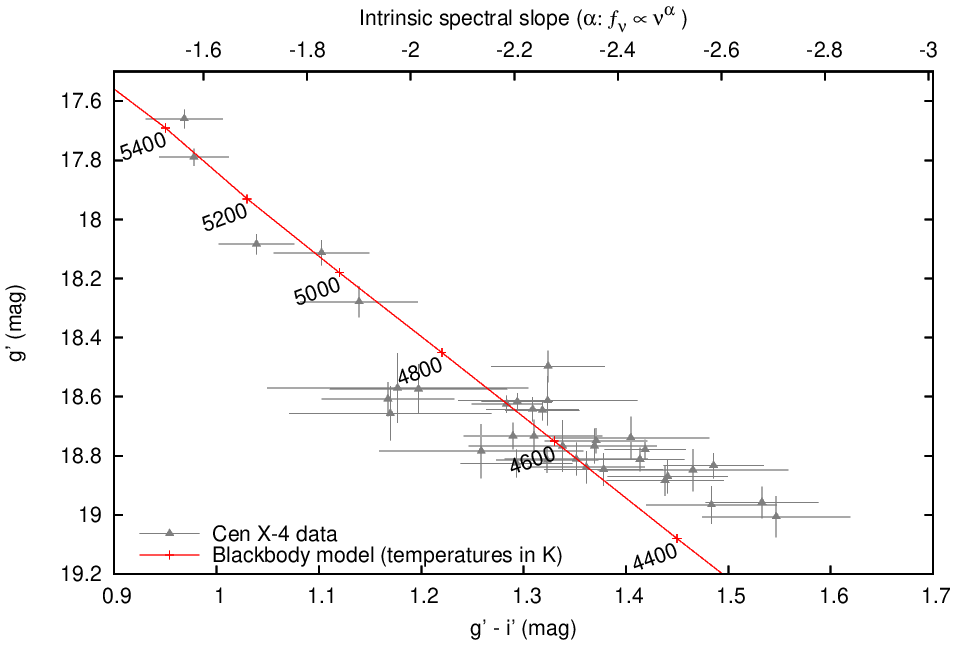}
\includegraphics[width=8.8cm]{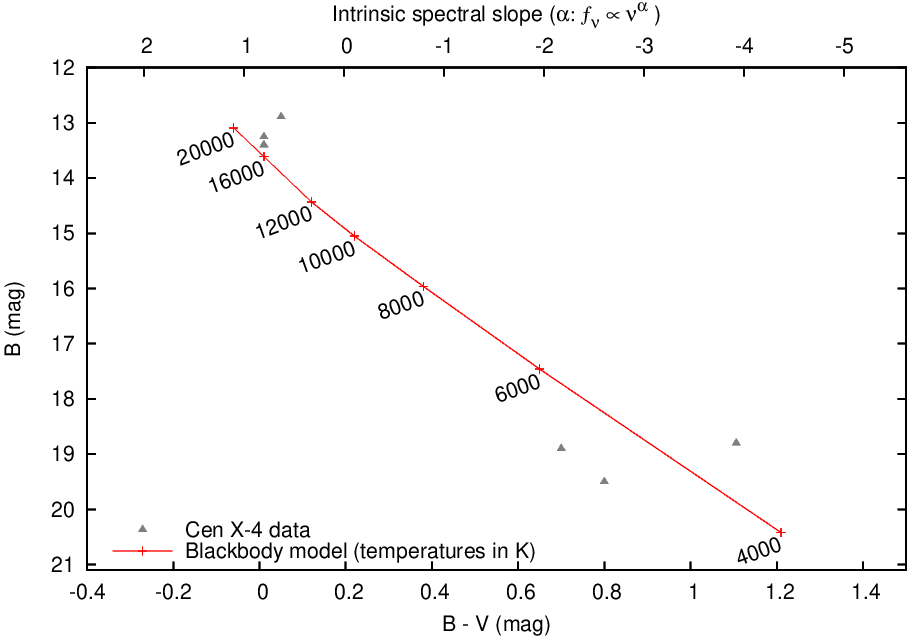}
\caption{\textit{Left}: Optical CMD ($g'$ vs. $g'-i'$) during the misfired outburst of Cen X-4. Bluer colors, corresponding to higher colour indices, are to the left, and redder colors (i.e. lower color index) to the right. The black body model is plotted with a red solid curve. Different temperature values are also highlighted close to the black body line. \textit{Right}: Optical CMD ($B$ vs. $B-V$) during the 1979 outburst of Cen X-4. The red solid line represents the black body model that best describes the data. Errors on the $B$-band points are not available in the literature, and therefore are not plotted. }
\label{cmd_figure}
\end{figure*}

\subsubsection{Spectral Energy Distribution}\label{Sec:SED}

In order to shed light on the nature of the misfired outburst, we built SEDs during the period of activity (using \textit{Swift}/UVOT and REM data obtained on January 4-5 2021) and during quiescence (using REM and LCO data acquired on May 22-23 2021). To do so, fluxes were de-reddened using the absorption coefficient $A_V=0.31\pm 0.16$ mag as reported in \citet{Russell2006}, and considering the relations of \citet{Cardelli1989} to evaluate the absorption coefficients at all wavelengths. Although from the light curves it is already clear that both disc and companion star contribute to the quiescent and flare emission of Cen X-4, we tried to model the two SEDs with a simple irradiated blackbody function, using the known parameters of the companion star (a $0.6\,R_{\odot}$ radius and a $0.35\, M_{\odot}$ mass; \citealt{Shahbaz1993}; \citealt{Torres2002}; \citealt{Shahbaz2014}). This of course constitutes a caveat, considering that no multi-temperature blackbody of the disc is added to the model; for a more precise determination of the accretion disc contribution, we refer the reader to Sec. \ref{Sec_CMD}. Since the least-squares fit is insensitive to the irradiation luminosity parameter, we fixed it to the measured X-ray luminosity of $L_{\rm X}=4.5\times 10^{32}\, \rm erg\,s^{-1}$ during quiescence \citep{Campana2004}, and to $L_{\rm X}=2.4\times 10^{33}\, \rm erg\,s^{-1}$ during the flare, as estimated from \textit{Swift} observations performed on 2021 Jan 4 (this work). The results are shown in Fig. \ref{fig:comparison_SED}. 

The fit of the quiescent SED obtained with REM and LCO in May 2021 (top panel of Fig. \ref{fig:comparison_SED}) gives comparable results to what was reported in \cite{Baglio2014}, with a blackbody temperature of $(4.43\pm0.01)\times 10^3\, \rm K$, consistent with a K5V-type star, as expected for Cen X-4 (\citealt{Shahbaz1993}; \citealt{Torres2002}). We note however that the NIR fluxes that we measure with our REM observations are lower with respect to the catalogued fluxes reported in 2MASS and published in \citealt{Baglio2014} (plotted as grey `x' in Fig. \ref{fig:comparison_SED}, upper panel). Considering that the 2MASS data were acquired in 2001, however, it is highly probable that the contribution from the accretion disc at optical and NIR frequencies  was different with respect to 2021 (also given the long-term trend observed in Fig. \ref{fig:long_monitoring}), thus explaining the discrepancy. 

In addition, we also plotted in Fig. \ref{fig:comparison_SED} (upper panel) the $V$, $R$, $i'$ fluxes obtained as the average emission from the lower envelope of the LCO long monitoring of Cen X-4 (Fig. \ref{fig:orbital_modulation}). These fluxes are the most constraining upper limits to the companion star contribution. The fluxes are lower with respect to the ones measured in quiescence with LCO by a factor of 1.7, 1.3 and 1.2 in $V$, $R$, $i'$ band, respectively. The fit of the three points with a black body gives a temperature of $(4.13\pm0.05)\times 10^3$ K, still consistent with a late type star.  

The fit of the flare SED (Fig. \ref{fig:comparison_SED}, bottom panel) with the irradiated star model gives a blackbody with a higher temperature, $T=(4.92\pm0.03)\times 10^3$ K.
The UV point during the flare cannot be described by this simplified irradiated star model, suggesting for an origin in the inner regions of the multi-temperature disc as it heats up, or, as reported in \citet{Bernardini2016} for observations during quiescence, a hot spot on the disc edge. Unfortunately it is not possible with our data to be conclusive on this.

We then subtracted the blackbody obtained by fitting the lower envelope fluxes from the flare SED. The result is shown in Fig. \ref{fig:subtracted_SED}.
The residual SED peaks below the $r'$ band; this suggests a residual component with temperature $< 5\times 10^3\, \rm K$ (according to the Wien displacement law, $T=b/\lambda$, where $b\sim 2897\, \rm \mu m\, K$ and $\lambda$ is the wavelength of the peak). It is therefore likely that we are observing the emission from a cold accretion disc, in the build-up for the start of an outburst (we note that according to the color-magnitude diagram shown in Fig. \ref{cmd_figure} the temperature of the disc at the beginning of the outburst is indeed $\sim 5\times 10^3 \, \rm K$). In Fig. \ref{fig:subtracted_SED}, the UV excess is also visible.

\subsubsection{Color-Magnitude diagram}\label{Sec_CMD}
We studied the color-magnitude diagram (CMD), $g'$ versus $g'-i'$, of Cen X-4 using LCO and REM data (Fig. \ref{cmd_figure}, left panel) obtained during the misfired outburst. Superimposed, we plot the blackbody model for an accretion disk, which depicts the evolution of a single-temperature, constant-area blackbody that heats up and cools down (for details: \citealt{Maitra2008}; \citealt{Russell2011}; \citealt{Zhang2019}). In the model, the color changes are determined by the different origins of the emission at optical frequencies: for low temperatures, the Rayleigh-Jeans blackbody tail; for high temperatures, the blackbody curved peak.
We note that this model assumes that the flux emitted by the source is all coming from the accretion disk, without any contribution from other sources (like the companion star), whereas the model depicted in Sec. \ref{Sec:SED} is assuming that the irradiated companion star is producing all the flux. Even though it is clear that both star and disk are contributing to the emission of Cen X-4, we consider these tests useful in order to shed light on the different contributions to the emission processes. 

We applied the disk model to Cen X-4 following \citet{Russell2011}, assuming an optical extinction of $A_{\rm V}=(0.31\pm0.16)$ mag \citep{Russell2006}, that is used to convert the color $g-i$ into an intrinsic spectral index (indicated on the top axis of Fig. \ref{cmd_figure}). The blackbody temperature depends on this color, while the normalization of the model depends on several different parameters; among these, the size of the blackbody and the distance to the source. Some of these parameters are uncertain so we varied the normalization of the model until a satisfactory approximation of the data was reached (see methods in \citealt{Maitra2008}, \citealt{Russell2011}, \citealt{Zhang2019}, \citealt{Baglio2020}).

The model approximates the data well, showing a trend that is consistent with a thermal blackbody. We therefore interpret this blackbody as that of the outer accretion disc (the surface area of the star is much smaller than that of the disc). Interestingly, the temperature of the disk remains low during the whole flare,
never exceeding $\sim 5400 \rm \,K$. This is also in agreement with the residual SED during the misfired outburst after the subtraction of the companion star contribution (Fig. \ref{fig:subtracted_SED}), where we observed a peak below the $r'$ band frequency, suggestive of a cold accretion disc ($T< 5\times 10^3 \, \rm K$).

We note that hydrogen is expected to be completely neutral below $5\times 10^3$ K (and completely ionised above $10^4$ K; \citealt{Lasota2001}).
It is therefore likely that the temperature required to start the heating wave in the disc, therefore kick-starting a full outburst, was never reached during the 2020/2021 activity. This condition is specific to this ``misfired'' outburst, as can be appreciated from the right panel of Fig. \ref{cmd_figure}, where the CMD ($B$ versus $B-V$ color) of Cen X-4 during the 1979 outburst (plus a few quiescent points) is shown, superimposed on a blackbody model which assumes the same normalization as during the 2020/2021 activity. In this outburst, the accretion disk reached and exceeded the temperature of 10000K, thus assuring the complete ionization of hydrogen in the disk, as required by the DIM to have a complete outburst. In particular, the brightest point in the CMD is found near the peak of the 1979 outburst. This shows that the data near the peak of the 1979 outburst are very close to the exact same model used to describe the 2020/2021 activity, thus reinforcing the idea that we witnessed a misfired outburst for Cen X-4 in 2021.

\subsubsection{Multi-wavelength correlation}

Another tool for disentangling the emission processes and for understanding the nature of the recent misfired outburst is the study of multi-wavelength correlations.

We studied the optical/X-ray correlation of the source during its flaring phase, using our LCO detections in the $i'$-band and quasi-simultaneous X-ray observations from {\it NICER} (taken within 1 hour). 
For the conversion of X-ray count rate to flux, we use a power law index of 1.7$\pm$0.3 \citep{Bernardini2013} and the same energy range (0.5-10 keV).

\begin{figure}
    \centering
    \includegraphics[width=8.9cm]{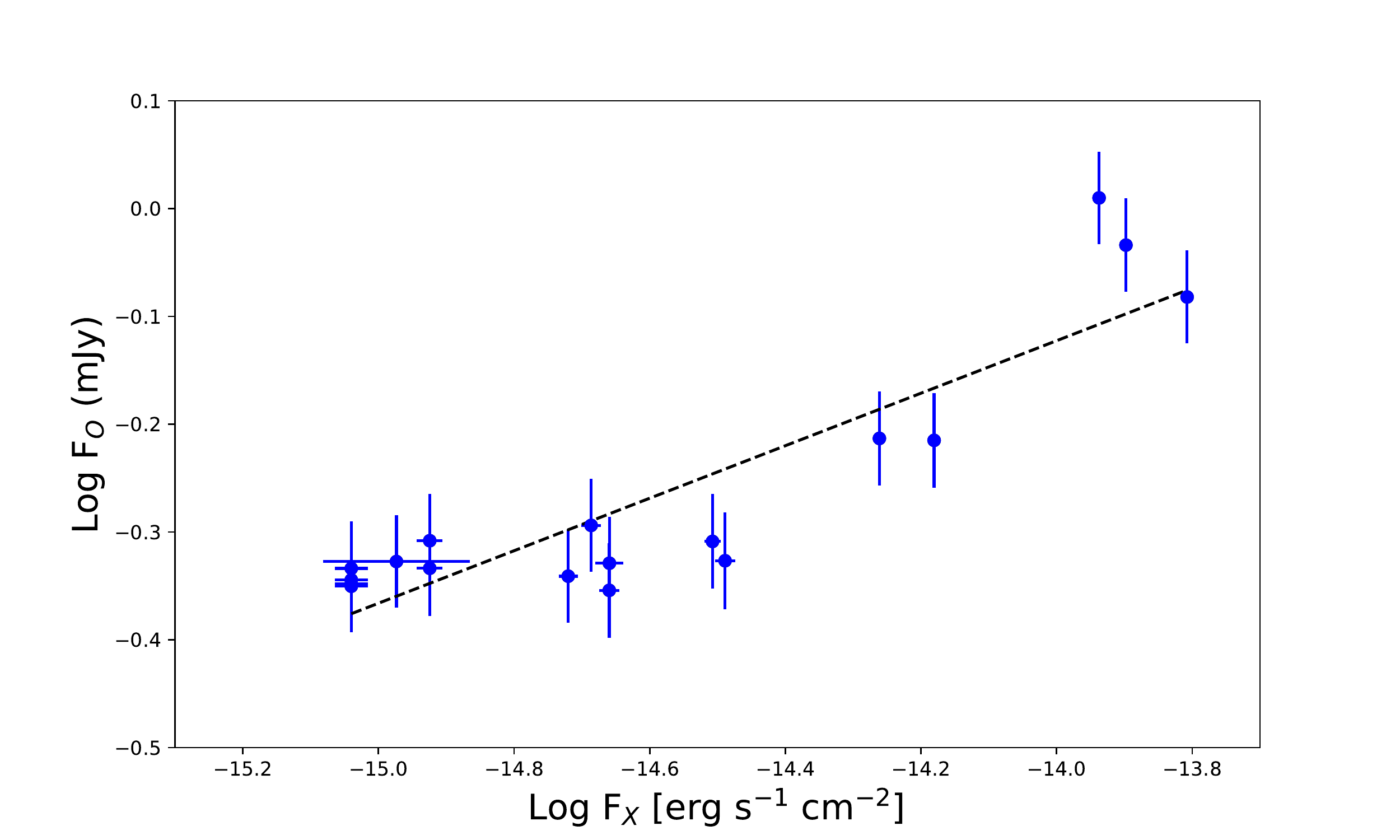}
    \caption{Optical/X-ray correlation during the recent flaring activity with quasi-simultaneous (within 1 hour) LCO $i$'-band optical data and {\it NICER} (0.5-10 keV) X-ray data.}
    \label{fig:correlation}
\end{figure}

We find a significant correlation between the optical and X-ray emission of the source during the flaring activity (with Pearson correlation coefficient = 0.89 and p-value = $8.2\times 10^{-7}$; see Fig. \ref{fig:correlation}). Previously, when the source was in quiescence, \citet{Cackett2013} found no significant correlation between the X-ray and simultaneous optical fluxes, while a positive correlation was observed between the X-ray flux and the simultaneous near-UV flux. Later, \citet{Bernardini2016} found evidence of optical ($V$-band), UV and X-ray correlation at quiescence on various timescales. The correlation slope found for outburst and quiescence had a slope of $\sim0.44$, showing that irradiation became important at high luminosities, but the slope is shallower than expected for irradiation near quiescence.

We fit the data during the misfired outburst using the orthogonal distance regression method of least squares, and find the slope of the optical/X-ray correlation to be 0.25$\pm$0.03, implying that irradiation is not playing a dominant role \citep[the expected slope for an irradiated disc is $\sim$ 0.5,][]{vanparadijs1994}. The observed slope would instead be more consistent with a viscously-heated accretion disc \citep[which can result in a slope $\sim$ 0.3, depending on the wavelength and on the compact object nature;][]{Russell2006}, or a combination of both \citep[e.g. $\sim$0.4 in GRS 1716-249,][]{Saikia1716}. For a viscously heated disc, a wavelength dependency of the optical/X-ray correlation slope has been observed for XRBs \citep{Russell2006}. In order to check for this, we studied the slope of the correlation using the four optical bands available during the misfired outburst ($i'$, $r'$, $V$, $g'$; Tab. \ref{tab:wavelengths}). Although the different values obtained were within the 1-sigma error range, we have found a slight trend of increasing slopes with increasing frequency (with values 0.25$\pm$0.03 for $i'$-band, 0.24$\pm$0.04 for $r'$-band, 0.30$\pm$0.04 for $V$-band and 0.35$\pm$0.06 for $g'$-band). This finding strengthens the argument that the optical emission originates from a viscously-heated accretion disc. From the optical/X-ray correlation coefficient, we can rule out the optical emission during the flaring activity to have an origin at the synchrotron jet \citep[for which the expected slope is much steeper, $\geq$ 0.7,][]{Russell2006}.

\subsection{DIM and inside-out outbursts}\label{sec:dim}

According to the DIM, when an XRB is in quiescence its accretion disc is cold and depleted. The mass transfer from the companion star, however, happening at low rates, replenishes the disc until the surface density at a certain annulus becomes sufficiently high to reach the critical density for which thermal equilibrium cannot be maintained. This makes the temperature of the ring increase over the hydrogen ionization temperature, and two different heating fronts begin to propagate inwards and outwards. 
Inside-out outbursts are most commonly observed in XRBs \citep[even though the heating fronts still propagate both ways;][]{Menou2000}, because they typically possess low accretion rates ($< 10^{16}$g/s; \citealt{vanparadijs1996}; \citealt{Smak1984}; \citealt{Menou1999}). Under these conditions, the accumulation time will be longer than the viscous time for diffusion, and matter will not be able to accumulate at the outer edge of the disc, resulting in an inward diffusion. 
Since the accretion rate decreases with radius, matter will then accumulate at a certain point, until it reaches the critical surface density for the thermal instability, triggering the inside-out outburst \citep{Lasota2001}. 
Inside-out outbursts typically propagate slowly \citep[][]{Menou2000}; in fact, the outward front encounters regions of higher density while propagating (and also the critical density will be higher for larger radii). In case the front is not transporting enough matter to raise the density at a certain radius above the critical density, the propagation will stall, and a cooling wave (propagating inwards) will be generated, which will prevent the outburst from occurring. Interestingly, a similar interpretation has also been given for the so-called Failed-Transition outbursts, i.e., outburst that do not reach the high/soft state \citep[][]{Alabarta2021}.
LMXBs are typically subject to strong irradiation, that is important to take into account in the DIM (\citealt{Dubus2001}; see also \citealt{TetarenkoB2018}, where actual data were used to test the DIM with irradiation). Irradiation has no effect on the structure of the heating front, but it is important to determine for how long the outward heating front will be able to propagate. In fact, with the propagation of the inside-out front, the mass accretion rate at the inner disc radius rises, therefore increasing the irradiation of the outer cold disc. As a consequence, the external disc is heated, which reduces the critical density needed to undergo the thermal instability, making the outward front propagation easier.

We estimate the mass accretion rate $\dot{M}$ of Cen X-4 during the misfired outburst using our X-ray monitoring. We integrated the count rates over the entire outburst and we converted it to flux, using the count rate to flux conversion factor obtained from the spectral fit in Sec. \ref{X-ray_sec}. We calculate the luminosity considering a distance of 1.2 kpc. Using $\dot{M}=L\, R_{\rm NS}/(G\, M_{\rm NS})$ (where $L$ is the X-ray luminosity, $R_{\rm NS}$ and $M_{\rm NS}$ are the typical radius and mass of a neutron star and $G$ is the gravitational constant; we note that we are assuming that all X-rays are due to accretion), and including an efficiency factor of $20\%$ in converting gravitational energy into luminosity \citep{Frank1987}, we estimate a mass accretion rate of $\sim1.5\times 10^{13}$g/s, that is considerably lower than the critical mass accretion rate that needs to be achieved in order to have outside-in outbursts\footnote{Even at the maximum X-ray flux during the misfired outburst, $\dot{M}$ only reached $5\times 10^{15}$g/s, $\sim 2$ orders of magnitude lower than the critical mass accretion rate.} \citep[considering Cen X-4 orbital parameters, $\dot{M}_{\rm crit}\sim 4\times 10^{17}$g/s;][]{Lasota2001}. Therefore, it is likely that an inside-out propagation front was ignited close to the inner radius of the accretion disc. At the time of the ignition, the temperature of the accretion disc according to the modeling of the CMD (Fig. \ref{cmd_figure}) was $\sim 5.4 \times 10^3\, \rm K$.
However, instead of an increasing temperature of the disc, what we observe in Fig. \ref{cmd_figure} is a temperature that decreases with time, from $\sim 5.4 \times 10^3\, \rm K$ to $\sim 4.4\times 10^3 \, \rm K$ and lower. In addition, the slope of the X-ray/optical correlation shows a scarce role of irradiation in the emission from the system, in agreement with previous studies performed during quiescence (see e.g. \citealt{Davanzo2006}), likely due to the very low mass accretion rate and to the large size of the system. 
It is possible that once the front started to propagate outwards, some irradiation was actually taking place, but the effect was low compared to all the other sources of emission in the optical (e.g. the companion star and the steady outer accretion disc, that emits in the optical). The overall optical emission would therefore dilute the effect of irradiation, explaining the shallow slope of the optical-X-ray correlation.
In addition, Cen X-4 is one of the XRBs with larger accretion discs known in the literature, due to its long orbital period ($\sim 15.1$ hr), which can explain the low level of irradiation to which the external accretion disc is exposed. 

We therefore conclude that the propagation of the outward front has likely stalled soon after the ignition due to the low mass accretion rate and low effect of the irradiation, the latter also linked to the known large size of the system. 
We note that the steep, short ($\sim 8-9$ days) decay phase after the misfired outburst peak is in agreement with the low level of irradiation that we observe in this work. In fact, the cooling front that is generated after the stall can only propagate if it finds a cold branch to fall onto \citep[][]{Lasota2001}; this is hampered by the effect of irradiation, that could keep the accretion disc hot, giving rise to the exponential and linear decay that is typically observed in strongly irradiated XRBs.

The factors which might have led to a misfired outburst are numerous. Among them, the size of the system surely makes a contribution, reducing the effect of irradiation and therefore facilitating the stall of the heating front propagation. We therefore predict that the larger the system is, the more likely it is for similar events to occur. 

Alternatively, as also suggested for the optical precursor to the 2019 outburst of the accreting millisecond X-ray pulsar SAX J1808.4-3658 \citep[][]{Goodwin2020}, the misfired outburst of Cen X-4 could have been caused by a local thermal instability at a radius close to the inner radius of the disc, where the density was close to the critical density at which the trigger of the full outburst could begin \citep[e.g.][Fig. 7]{Menou2000}. This interpretation could work for Cen X-4, considering that the temperature in the disc has always remained below $6\times 10^3$ K (i.e. the temperature of hydrogen ionization).
Had the full outburst actually started for Cen X-4, the misfired outburst described in this work would therefore have been its precursor.

\section{Conclusions}
In this work we report on the long term optical monitoring of the neutron star low mass X-ray binary Cen X-4 during the past 13.5 years. The source spent the majority of this time in quiescence; the ellipsoidal modulation due to the companion star emission can be isolated, together with several short-timescale variations in all optical bands, likely due to activity in the accretion disc. Once the flares and the ellipsoidal modulation from the star are subtracted, the residual flux shows a linear downward trend spanning $\sim 3000$ days, followed by an upward trend for about 1000 days, $\sim7$ times steeper than the downward one. In the case of the black hole X-ray binary V404 Cyg \citep{Bernardini2016_precursor}, a similar upward trend of the flux preceded the start of an outburst in 2015. However, although a significant brightening was observed at the beginning of 2021 at all wavelengths (NIR--X-rays), a proper outburst was not triggered in the case of Cen X-4, which returned to quiescence a few weeks after the start of this enhanced activity. We term this behaviour as a ``misfired outburst'', because the temperature required to ionize hydrogen and initiate the outburst, was not reached.
The modeling of the color magnitude diagram during the misfired outburst with a single-temperature black body shows an accretion disc with temperatures below $5.4\times 10^3\, \rm K$; this result is in agreement with the residual spectral energy distribution, after the subtraction of the contribution from the companion star, and suggests that the accretion disc never reached the temperature that is required to ionize hydrogen (in contrast to what happened during the 1979 full outburst of the source, when, according to our model, the accretion disc reached temperatures of $\sim 2\times 10^4\, \rm K$, where hydrogen is typically completely ionized).

A possible interpretation is that an inside-out type outburst was initiated. Inside-out outbursts typically propagate slowly, because the heating front meets regions of higher density while propagating outwards. If the front is not transporting enough matter, it will stall unless irradiation is strong enough to heat the external disc, therefore decreasing the surface density and facilitating the propagation.
However, irradiation is scarce in Cen X-4. In fact, the optical/X-ray correlation during the misfired outburst has a shallow slope, inconsistent with a strongly irradiated disc; moreover, it was already reported in the past \citep[see e.g.][]{Davanzo2006} that the effects of irradiation are low in Cen X-4, consistent with the large size of the system.
It is therefore likely that the heating front was halted soon after its ignition, with a consequent production of an opposite cooling front, which switched off the outburst.
Alternatively, the observed activity could be the result of a local thermal-viscous instability in the disc, where temperatures increased without however reaching (and overcoming) the temperature for hydrogen ionization.
The optical monitoring of Cen X-4 is still ongoing, and will show whether a new misfired or full outburst might happen in the future, thus shedding further light on the possible mechanisms preventing a complete outburst to be triggered.

\acknowledgments
We thank the anonymous referee for useful comments and suggestions. 
This research has made use of data and/or software provided by the High Energy Astrophysics Science Archive Research Center (HEASARC), which is a service of the Astrophysics Science Division at NASA/GSFC.
This work is also based on observations made with the REM Telescope, INAF Chile, and makes use of observations performed with the Las Cumbres Observatory network of telescopes.
DMR and DMB acknowledge the support of the NYU Abu Dhabi Research Enhancement Fund under grant RE124.
J.H. acknowledges support for this work from the {\it NICER} Guest Investigator program under NASA grant 80NSSC21K0662. We thank the {\it Swift} and {\it NICER} teams for rapidly approving, scheduling, and performing the X-ray observations.
SC and PDA acknowledge support from ASI grant I/004/11/5.
JvdE is supported by a Junior Research Fellowship awarded by St. Hilda's College, Oxford.
NM acknowledges the ASI financial/programmatic support via
the ASI-INAF agreement n. 2017-14-H.0 and the 'INAF Mainstream' project on the same subject.
TMD acknowledges support from the Spanish ministry of science under grant EUR2021-122010. TMD acknowledges support from the Consejeria de Economia, Conocimiento y Empleo del Gobierno de Canarias and the European Regional Development Fund under grant ProID2020-010104. 

\bibliography{bibliography}
\bibliographystyle{aasjournal}



\end{document}